\documentclass[12pt]{article}

\usepackage{graphicx,epstopdf}
\usepackage{amsmath}
\usepackage{epsf}

\def\be{\begin{equation}}
\def\ee{\end{equation}}
\def\ba{\begin{eqnarray}}
\def\ea{\end{eqnarray}}
\def\ge{\mathrel{\raise.3ex\hbox{$>$\kern-.75em\lower1ex\hbox{$\sim$}}}}
\def\la{\mathrel{\raise.3ex\hbox{$<$\kern-.75em\lower1ex\hbox{$\sim$}}}}

\def\simgt{\mathrel{\raise.3ex\hbox{$>$\kern-.75em\lower1ex\hbox{$\sim$}}}}
\def\simlt{\mathrel{\raise.3ex\hbox{$<$\kern-.75em\lower1ex\hbox{$\sim$}}}}

\newcommand{\bi}[1]{\bibitem{#1}}
\newcommand{\fr}[2]{\frac{#1}{#2}}

\newcommand{\nc}{\newcommand}

\nc{\gone}{\bar g_{\pi NN}^{(1)}}
\nc{\gzero}{\bar g_{\pi NN}^{(0)}}
\nc{\al}{\alpha}
\nc{\ga}{\gamma}
\nc{\de}{\delta}
\nc{\ep}{\epsilon}
\nc{\ze}{\zeta}
\nc{\et}{\eta}
\nc{\Th}{\Theta}
\nc{\ka}{\kappa}
\nc{\rh}{\rho}
\nc{\si}{\sigma}
\nc{\ta}{\tau}
\nc{\up}{\upsilon}
\nc{\ph}{\phi}
\nc{\ch}{\chi}
\nc{\ps}{\psi}
\nc{\om}{\omega}
\nc{\Ga}{\Gamma}
\nc{\De}{\Delta}
\nc{\La}{\Lambda}
\nc{\Si}{\Sigma}
\nc{\Up}{\Upsilon}
\nc{\Ph}{\Phi}
\nc{\Ps}{\Psi}
\nc{\Om}{\Omega}
\nc{\ptl}{\partial}
\nc{\del}{\nabla}
\nc{\ov}{\overline}
\nc{\newcaption}[1]{\centerline{\parbox{15cm}{\caption{#1}}}}

\nc{\hef}{$^4$He}
\nc{\het}{$^3$He}
\nc{\lisx}{$^6$Li}
\nc{\lisv}{$^7$Li}
\nc{\bes}{$^7$Be}
\nc{\beet}{$^8$Be}
\nc{\hefm}{{\rm ^4He}}
\nc{\hetm}{{\rm ^3He}}
\nc{\lisxm}{{\rm ^6Li}}
\nc{\lisvm}{{\rm ^7Li}}
\nc{\besm}{{\rm ^7Be}}
\nc{\beetm}{{\rm ^8Be}}
\nc{\bs}{(N$X^-$)}
\nc{\xm}{$X^-$}
\nc{\xp}{$X^+$}
\nc{\xz}{$X^0$}
\nc{\bex}{(\bes\xm)}
\nc{\bexm}{(\besm X^-)}
\begin{document}

\begin{titlepage}
\rightline{UVIC--TH--07/02}
\rightline{hep-ph/yymmnnn}
\rightline{March 2007}
\begin{center}

\vspace{0.5cm}

\large {\bf Primordial Lithium Abundance in Catalyzed Big Bang Nucleosynthesis}
\vspace*{5mm}
\normalsize

{\bf Chris Bird$^{\,(a)}$, Kristen Koopmans$^{\,(a)}$, and Maxim Pospelov$^{\,(a,b)}$}

\smallskip
\medskip

{a. \it Department of Physics and Astronomy,  University of Victoria,
Victoria, BC, V8P 1A1 Canada}

{b. \it Perimeter Institute for Theoretical Physics, Waterloo,
Ontario N2J 2W9, Canada}

\smallskip
\end{center}
\vskip0.6in

\begin{abstract}
There exists a well known problem with the \lisv+\bes\ abundance predicted by standard big bang nucleosynthesis being larger than the
value observed in population II stars. The catalysis of big bang nucleosynthesis by  metastable, $\tau_X \ge 10^3$ sec,
charged particles $X^-$ is capable of suppressing the primordial \lisv+\bes\, abundance and
making it consistent with the observations. 
We show that to produce the correct abundance, this mechanism of suppression places a requirement on the initial abundance of $X^-$ at temperatures 
of $4\times 10^8$ K to be on the order of or 
larger than $0.02$ per baryon, which is within the natural range of abundances 
in models with metastable electroweak-scale particles. The suppression of \lisv+\bes\, is triggered by the formation of
\bex\, compound nuclei, with fast depletion of 
their abundances by catalyzed proton reactions, and in some models by direct capture of \xm\ on \bes. 
The combination of \lisv+\bes\, and \lisx\ constraints 
favours the window of lifetimes, $1000~{\rm s}~\la\tau_X\leq 2000~{\rm s}$.

\end{abstract}
\vspace*{2mm}

\end{titlepage}

\section{Introduction}
\par
Big Bang Nucleosynthesis (BBN) is the process of light element formation during the early stages of
cosmological expansion. Theoretical predictions of elemental abundances are based on known 
physics such as the nuclear reactions of light elements, 
general relativity and the Standard Model (SM) of particles physics. If the observations of 
hydrogen, helium and lithium abundances in the present are capable of determining their primordial values, the consistency 
of the entire theoretical framework can be tested. In recent years these tests acquired particular 
sharpness, as the only free parameter, the ratio of baryons to photons, is now well-measured 
through the anisotropy of the cosmic microwave background \cite{WMAP}. 
\par
Perhaps the most exciting prospect of studying the primordial abundances is the possibility 
of testing the combination of Standard Model and general relativity, which we call SBBN, 
or standard BBN. To this end it is important to understand how 
the non-standard physics can affect the outcome of nuclear reactions (see {\em e.g.} \cite{Sarkar} for 
a review). Schematically, the BBN equations
can be represented as
\begin{eqnarray}
\frac{dY_i}{dt}=-H(T)T \frac{d Y_i}{ dT } = \sum (\Gamma_{ij}Y_j+ \Gamma_{ikl}  Y_k Y_l+...);
~~~~~{\rm Energy~of~reactants}\sim T\la {\rm MeV}\nonumber\\
H(T) = {\rm const}\times  N_*^{1/2} \fr{T^2}{M_{\rm Pl}},~~
{\rm where} ~ N_* = N_{boson} +\fr78N_{fermion}.~~~~~~~~~~~~~~~~~~
\label{framework}
\end{eqnarray}
In this formula, $Y_i$ are the abundances of light elements, $\Gamma_{ij...}$ are the generalized
(positive or negative) rates for creation or destruction of element $i$ with participation of  $j,k...$, $H(T)$ is the Hubble 
expansion rate, $M_{\rm Pl}$ is the Planck constant, and 
$N_*$ is the number of effective degrees of freedom comprised of fermionic 
and bosonic fields. The radiation-domination expression for $H(t) = 1/(2t)$ is used.
There are several ways in which non-standard cosmological and/or particle physics 
can affect the freeze-out abundances $Y_i(T\to0)$ of light elements. To date the following 
generic possibilities have been identified: 

\begin{enumerate}

\item {\em Timing} of the reactions can be changed, for example by models that have 
additional contributions to $H(T)$ (See Ref. \cite{Sarkar} and references therein). 
Such contributions may come from the additional 
thermally excited relativistic degrees of freedom (historically often referred to as 
``extra neutrino" species), or from any other forms of energy that contributes to the total 
energy density at a significant level, such as ``tracker" scalar fields \cite{Ed}. 

\item {\em Non-thermal components} to nuclear reactions can be introduced by injection
of energy during or after the BBN \cite{metastable}. For example, unstable or annihilating heavy particles
can cause such injection, leading for example to the break-up of \hef\, into D or \het\,,
or to the synthesis of lithium via out-of-equilibrium processes involving energetic secondary projectiles. 
The most important parameters in these types of models are the amount of energy injected
at $t\sim$ lifetime, and the fraction of energy injected into hadronic degrees of 
freedom. 

\item {\em New catalyzed thermal channels} $\Gamma_{ij..}$ can be opened up if the heavy particle
physics remnants have the ability to bind to nuclei during the BBN. The consequences of the 
catalyzed BBN (CBBN) 
scenario with charged relics which are present at the time of the BBN, 
and the corresponding catalytic enhancement of \lisx\ by several orders of magnitude, was explicitly demonstrated 
in the recent paper \cite{CBBN}. The main parameters that determine the strength of catalysis are 
the initial number densities of such particles and their lifetimes. 

\end{enumerate}

Other notable modifications to the standard scenario include an impact from possible inhomogeneity 
of the Universe at the time of BBN  \cite{inhom}, and even more contrived options such as 
 time dependent couplings \cite{alpha}. The latter again affects the BBN predictions mostly via 
changing the timing of such processes as the $n/p$ freeze-out and deuterium formation. 

\par
In this paper we assess the ability of CBBN to reduce the amount of primordial  
\lisv. This question has gained importance recently, as the amount of 
\lisv\, predicted by  SBBN is about a factor of $2-3$ larger than the amounts of 
\lisv\, observed in the stellar atmospheres at low metallicities. This discrepancy is 
difficult to attribute to the nuclear rate uncertainties in 
the context of the standard scenario \cite{french,belg}. It is entirely possible, however, that unaccounted 
depletion of \lisv\, in stars can be responsible for the overall  depletion of
primordial \lisv. We refer the reader to the recent discussion in astrophysical 
literature \cite{astro}. Regardless of whether the primordial 
lithium overabundance issue can be resolved by careful analysis of realistic stellar models, 
one should investigate the feasibility of other options for the reduction of primordial 
lithium, such as by new models of particle physics. It has been shown recently that the 
reduction of \lisv\, from a large energy release is possible if the metastable particles
decay at around $30$ keV temperature ($\tau_X \sim (1-3)\times 1000$ sec) and 
have a significant hadronic fraction in their decay
products \cite{jj}. Another possibility based purely on the catalysis of thermal nuclear reactions 
was pointed out in \cite{CBBN}, where a similar range of lifetimes has been 
suggested. Besides Ref. \cite{CBBN}, the cosmological consequences from the bound states of nuclei with charged 
particles were considered in \cite{Khlopov,KT,KR,Cy} (see also the discussion in the earlier work
\cite{Dimopoulos:1989hk}).
\par
The efficiency of the catalytic reduction of \lisv, or more precisely of \lisv+\bes, depends
crucially on the rate of the \bex\, bound state formation , as more than $90\%$ of predicted 
lithium at $\eta_b=\eta_b^{WMAP}$ comes in the form of \bes. Once \bex\, is formed, a new set of 
reaction channels opens up \cite{CBBN}, and this paper considers them in detail. 
These channels include an enhancement in $p$-destruction of \bes, 
enhanced internal capture of \xm\, and \bes, as well as other channels. 
The \bex\, bound state then serves as a ``bottleneck" for a potential solution to lithium 
problem in the CBBN framework. It is therefore very important to calculate the 
efficiency of the \bex\, formation as accurately as possible, as the naive hydrogen-like
formula employed in some studies may give a poor approximation to the correct answer. 
Having determined the catalyzed nuclear rates, and the rate of the 
bound state formation, we solve the resulting BBN equations in the 
lithium chain numerically  for some representative values of the parameter 
space and determine in which cases CBBN provides a noticeable reduction of 
\bes+\lisv, while keeping \lisx\, within observational bounds. 
\par
It is important to stress that we try to keep our investigation of 
CBBN processes as model independent as possible. The two main parameters that enter in the 
CBBN reactions are the lifetime and the initial abundance of the \xm\ particles, which we 
choose to normalize on the baryon number density,
\be
\fr{n_X}{n_b} =  Y_X  \exp(-t/\tau_X)= Y_X\exp(-178{\rm s}/(\tau_XT_9^{2})).
\ee
In this formula, $T_9$ is the temperature in units of $10^9$ K, $\tau_X$ is the lifetime, and $Y_X$ is the initial 
abundance at $t\ll\tau_X$. 
Besides the existence of $\{Y_X,\tau_X\}$ parameter space, 
one has to recognize the existence of two generic classes 
of models that can affect the CBBN reactions. These are models of  
type I and type II \cite{CBBN}, which have different ways of achieving the longevity of \xm. In type I models 
the long lifetime is due to the weakness of the couplings (e.g. models with the gravitino as the lightest supersymmetric particle), 
while in type II it is due to the 
small, $O$(MeV),  energy release in the transition between \xm\ and \xz\ 
while the couplings are assumed to be of normal size. Type I models are 
necessarily accompanied by large energy release, and thus possible impact on BBN can be 
twofold: due to the catalysis of thermal nuclear reactions and due to non-thermal processes. 
In this paper we do not treat non-thermal effects, as they tend to be very model-specific.  

\section{Recombination of \bes\, and \xm}

\subsection{ Properties of the bound states} The properties of the bound states of 
\bes\, and \xm\, are very important because they are essential for
determining the temperature and probability of recombination. Using the Gaussian charge distribution 
within the \bes\, nucleus, we determine the following energy of the ground state 
relative to \bes+\xm\ continuum:
\begin{eqnarray}
\label{Eb}
E^{(0)}_g = -2785~{\rm keV};~~a_B=1.03{\rm fm}~~~~ {\rm for} ~~~~ \langle r^2 \rangle_{\rm Be} = 0 \\
E_g = -1330\pm 20 ~{\rm keV}~~~~ {\rm for} ~~~~ (\langle r^2 \rangle_{\rm Be})^{1/2} = 2.50\pm 0.04{\rm fm} \nonumber
\end{eqnarray}
In this formula, the first line is a naive Bohr-like formula for a point-like nucleus,
which, of course, gets a significant correction from the realistic values 
of the charge radius. These values can be derived from experimentally measured values for
the nucleon radius of \bes\, \cite{radii}. In the remainder of the paper we adopt the central value from 
the second line of (\ref{Eb}). We have checked that both the variational calculation and the 
numerical solution to the Schr\"odinger equation produce identical answers. The radial profiles 
for the wave function are plotted in Figure 1.

\begin{figure}[htbp]
\centerline{\includegraphics[width=11 cm]{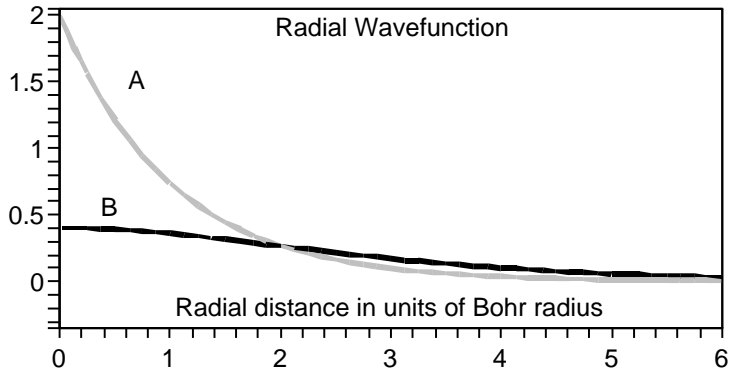}}
 \caption{\footnotesize Normalized radial wavefunction for the bound state \bex\, as a function of 
distance in units of $a_B = 1.03$ fm. A. A
hydrogen-like  profile for a ``point-like" \bes. B. 
Realistic wavefunction for the Gaussian charge distribution inside \bes.}
\label{f1} 
\end{figure}

Information about the ground state alone is insufficient 
for calculating the realistic recombination rate. 
The internal structure of the \bes\, nucleus must also be taken into account. In particular, the 
first excited state of beryllium is close to the ground state and cannot be neglected,
\be
|E_{\besm_{3/2}} - E_{\besm_{1/2}}| = 429~{\rm keV}
\ee
Indeed an excited form of \bes, which has total angular momentum $I=1/2$ (as opposed to 
the ground state with $I=3/2$) can also bind to \xm. Some of these metastable bound states
have total energies just above the \bes+\xm continuum threshold. This opens the possibility of 
a resonant transition into a bound state, and therefore such states have to 
be taken into account separately. The three most important bound states identified in this way 
belong to the principal quantum number $n=3$ 
\begin{eqnarray}
(^7{\rm Be}_{1/2} X^-), ~~ n=3,~~l=0,~~E_R=(-239+429){\rm keV= 190 keV};\nonumber\\
\phantom{(^7{\rm Be}_{1/2} X^-),}~~ n=3,~~l=1,~~E_R=(-290+429){\rm keV= 140 keV};\\
\phantom{(^7{\rm Be}_{1/2} X^-),}~~ n=3,~~l=2,~~E_R=(-308+429){\rm keV= 121 keV}.\nonumber
\end{eqnarray}
In these formulae, $E_R$ is the energy level of the state relative to the
$\besm_{3/2}$+\xm\, continuum. 

It is also of interest to quote the position of $n=2$ resonances:
\begin{eqnarray}
(^7{\rm Be}_{1/2} X^-), ~~ n=2,~~l=0,~~E_R=(-478+429){\rm keV= -49 keV};\nonumber\\
\phantom{(^7{\rm Be}_{1/2} X^-),}~~ n=2,~~l=1,~~E_R=(-636+429){\rm keV= -206 keV}.
\end{eqnarray}
Unlike the hydrogen-like case,  the $2l$ are not exactly degenerate, which allows for a significant 
one-photon $E1$ transition between them, $\Gamma_{2s\to2p} \simeq 0.5 eV$, 
and is quite comparable to the electromagnetic widths of $n=3$ states. 
One can also notice that the $2s$ state is very close to the 
threshold, which is potentially very important for the calculation of the capture rate.  
Moreover, there are number of additional effects that may lead to a few tens of keV upward shift 
of this level. Such effects include the correction for the finite mass of $X^-$, a perhaps larger charge radius 
of the excited state, and correction for nuclear polarizability. The precision in our calculations 
is certainly limited, and only the dedicated many-body nuclear studies can reach better than $O(50{\rm keV})$ 
determination of the energy levels. We therefore conclude that the $2s$ state may be indeed at the threshold
or just above $E=0$.

\subsection{ Resonant and non-resonant recombination}
The process of recombination between \bes\,  and \xm\, may seem to be relatively simple to calculate. 
For the continuum spectrum wave functions we can 
use the approximation of $E\simeq 0$ to good accuracy, as 
the characteristic energies are on the order of the temperature, which is significantly smaller 
than the binding energy, $T/ E_g\sim 0.03$. We find these wave functions by solving the 
corresponding Schr\"odinger equation numerically. The process of recombination can occur 
via two mechanisms involving resonant and non-resonant capture. 
\par
The non-resonant recombination is given by the processes
\ba
\label{nonres}
\besm_{3/2}+X^-\to (\besm_{3/2}X^-,n\geq1)+\gamma\to (\besm X^-)+k\gamma,\\
\besm_{3/2}+X^-\to (\besm_{1/2}X^-,n=1,2)+\gamma\to (\besm X^-)+k\gamma, \nonumber
\ea
where $k$ is a typical number of emitted photons ranging from 1 to 3. 
Among the non-resonant processes the capture 
of \bes $_{3/2}$ directly to the ground state has the largest cross section, closely followed by the 
capture to $n=2,l=0$ level. The cross section of photorecombination differs very significantly
from the ``Hydrogen-like" formula. In fact, it is about one order of magnitude smaller than 
the naive formula, with most of the suppression coming from the $\omega^3$ factor , which is 
a factor of $\sim$8 smaller in the realistic case (\ref{Eb}). 

The resonant capture occurs through the following chain of transitions:
\be 
\label{res}
\besm_{3/2}+X^-\to (\besm_{1/2}X^-,n\geq3)\to 
(\besm_{1/2}X^-,n=1)+k\gamma\to (\besm X^-)+(k+1)\gamma,
\ee
To estimate the cross section in this case, one has to find the widths of the 
corresponding metastable states. It turns out that the entrance width of the 
process $\besm_{3/2}+X^-\to (\besm_{1/2}X^-, n=3)$ is due to a quadrupole nuclear transition
in the electric field of \xm. If the distance $r$ between \xm\ particle and 
\bes\ is large, the interaction takes the form of  $V_{int} = \fr{1}{6}Q^{nucl}_{ij}\nabla_i \nabla_j (\alpha/r)$. 
using the existing experimental information on the quadrupole 
matrix elements in \bes\ system, we estimate matrix elements of this interaction 
between the initial wave function of the free
$\besm_{3/2}$, and the intermediate wave function of the $3l$ bound states of $\besm_{1/2}$,
to conclude that $\Gamma^{in}_{Q}$
is comparable to 1 keV. Given that the electromagnetic decay widths of $3l$ states,
whose calculations are much easier,  are all on the order of $\Gamma_{\gamma}^{out} \sim$ eV, we arrive 
at the following hierarchy of the "in", "out" widths and the temperature,
\be
\Gamma_\gamma^{out} \ll \Gamma_{Q}^{in}\ll T.
\ee 
This warrants the narrow resonance approximation in the Breit-Wigner formula,
and we use 
\be \sigma_R \sim g\sigma_{geom}\times \pi \Gamma_\gamma \delta(E-E_R)
, \label{BW}
 \ee
 where $g$ is the corresponding multiplicity factor, and $\sigma_{geom}$ is the 
 geometric cross section. Notice that in this approximation the rather uncertain 
 value of $\Gamma_{Q}^{in}$ does not enter into the formula for the cross section.
 
\begin{figure}[htbp]
\centerline{\includegraphics[width=11 cm]{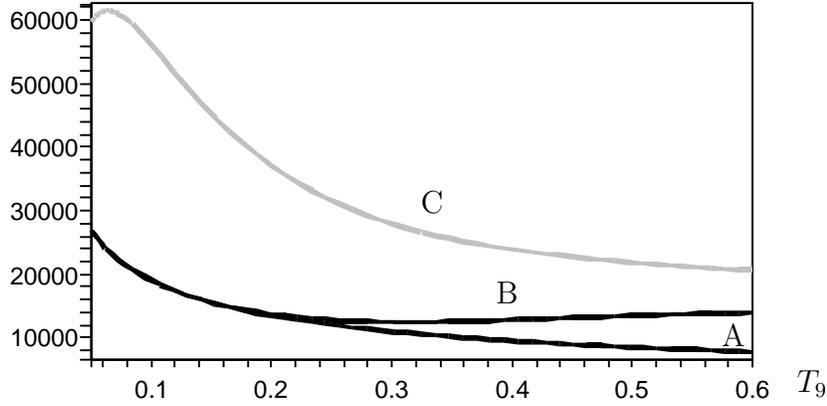}}
 \caption{\footnotesize Recombination rate
$\langle \sigma_{rec}v\rangle$ for \bes\, and \xm\, in 
astrophysical units. A: Nonresonant contribution. B: Total recombination rate
including $3l$ resonances. C: total recombination rate, including 
the $2s$ level, in the assumption of $E_R \sim 10$ keV. 
Given possible $O(50)$ keV uncertainty in the position of the 2s level, the whole area between 
curves A and C is representative of the nuclear uncertainty in the recombination rate.}

\vspace{-5.5cm} \hspace{8cm} C

\vspace{0.7cm} \hspace{9cm} B

\vspace{0.05cm} \hspace{12cm}  A

\vspace{0.1cm} \hspace{13cm}  $T_9$

\label{f2_mod} 

\vspace{3cm}
\end{figure}

 After performing several quite tedious but straightforward calculations, we arrive at the 
 total recombination cross section, averaged over the thermal distribution of 
 \xm\, and \bes:
 \be
 \label{sigma_tot}
 \langle \sigma_{rec} v\rangle = 
 \frac{6\times 10^3}{T_9^{1/2}}+\frac{1.9\times 10^4}{T_9^{3/2}}\exp(-1.40/T_9)+
\frac{1.5\times 10^4}{T_9^{3/2}}\exp(-1.62/T_9).
\ee
Here $T_9$ is temperature in  $10^9$ Kelvins, and the rate is 
expressed in conventional astrophysical units of $N_A^{-1}$cm$^3$s$^{-1}$mol$^{-1}$.
The first term in (\ref{sigma_tot}) corresponds to the non-resonant processes of (\ref{nonres}),
while the second and third terms are the resonant pieces with internal excitations of \bes\, and
initial capture on $n=3,l=2$ and $n=3, l=1$ levels. The relative size of the 
two effects can be seen in Figure \ref{f2_mod}, where the total recombination 
cross section, and the non-resonant piece are plotted separately. One can see that the resonant effects
from $3l$ levels may provide a sizable (up to 50\%) contribution 
in the most important domain of temperatures, $T_9 \sim 0.4$. 

So far the possible contribution of $n=2$ states has been ignored. There is, however, an ample 
chance of $2s$ state giving a large contribution to the recombination cross section via the following chain,
\be 
\label{res2s}
\besm_{3/2}+X^-\to (\besm_{1/2}X^-,n=2,l=0)\to 
(\besm_{1/2}X^-,n=2,l=1)+\gamma\to (\besm X^-)+3\gamma.
\ee
As was argued in the previous subsection, only a dedicated nuclear calculation can 
exactly determine the position of this resonance, which consequently bring a sizable
nuclear uncertainty into the calculation of recombination cross section. In what follows we assume 
two extreme situations. One is when the $2s$ level is sufficiently below the threshold so that it does not 
contribute significantly to the non-resonant part of the rate, in which case Eq. (\ref{sigma_tot}) will be adopted.
Another case, is when we assume that the $2s$ level is pushed upward to the resonant energy of $+10$ keV. In this case 
we estimate that the total recombination rate from (\ref{sigma_tot}) receives an additional contribution 
of 
\be
\label{addon}
\langle \sigma_{rec} v\rangle \to \langle \sigma_{rec} v\rangle  + \frac{4\times 10^3}{T_9^{3/2}}\exp(-0.12/T_9),
\ee
and the total result of (\ref{addon}) can be viewed as the most "optimistic" estimate of the recombination 
rate. 
This contribution is controlled directly by $\Gamma_{2s\to2p}$. If indeed the resonant energy is around 10 keV or so, 
then for all practical purposes the exponent in (\ref{addon}) can be taken as $\sim 1$. As one can see from 
Fig. \ref{f2_mod}, the contribution of $2s$ state is capable of enhancing the recombination rate by a factor of a few, 
thus signaling a significant nuclear uncertainty.

The photodisintegration rate of \bes\, is one-to-one related to the direct rate via the detailed balance 
relation \cite{reactions}:
\be
\Gamma_{photo} = \int_{E>|E_g|}  \sigma_{photo} dn_\gamma(E)= \langle \sigma_{rec} v\rangle \times
\left(\fr{m_{\rm Be} T}{2\pi}\right)^{3/2}\exp(-|E_g|/T).
\label{out}
\ee
Numerical expression for the photoionization rate is given by
\be
\fr{\Gamma_{photo}}{n_\gamma}= \langle \sigma_{rec} v\rangle \times \fr{5.5\times10^6}{T_9^{3/2}}\exp(-15.42/T_9) ,
\ee
where $n_\gamma(T)= 0.24T^3$ is the total number density of photons.

Combining the two rates, (\ref{sigma_tot}) and (\ref{out}), one arrives at the 
recombination equation for the abundance of bound states 
\bex\, relative to the total number of \bes\, nuclei,
\be
-HT\fr{dY_{BS}}{dT}=n_X(1-Y_{BS})\langle \sigma_{rec} v\rangle - \Gamma_{photo}  Y_{BS} ,
\ee
where we also assume that $n_X\gg n_{\besm}$ and took the limit of large $\tau_X$. 
One can easily see that in the limit $ \Gamma_{rec};~\Gamma_{photo} \gg H$ this equation has an 
attractor Saha-type solution,
\be
Y_{BS} \to Y_{BS}^{\rm Saha} = \left[1+(m_{\rm Be} T/2\pi)^{3/2}n_X^{-1}\exp(-|E_g|/T)\right]^{-1},
\ee
which corresponds exactly to the case of chemical equilibrium.

\begin{figure}[htbp]
\centerline{\includegraphics[width=11 cm]{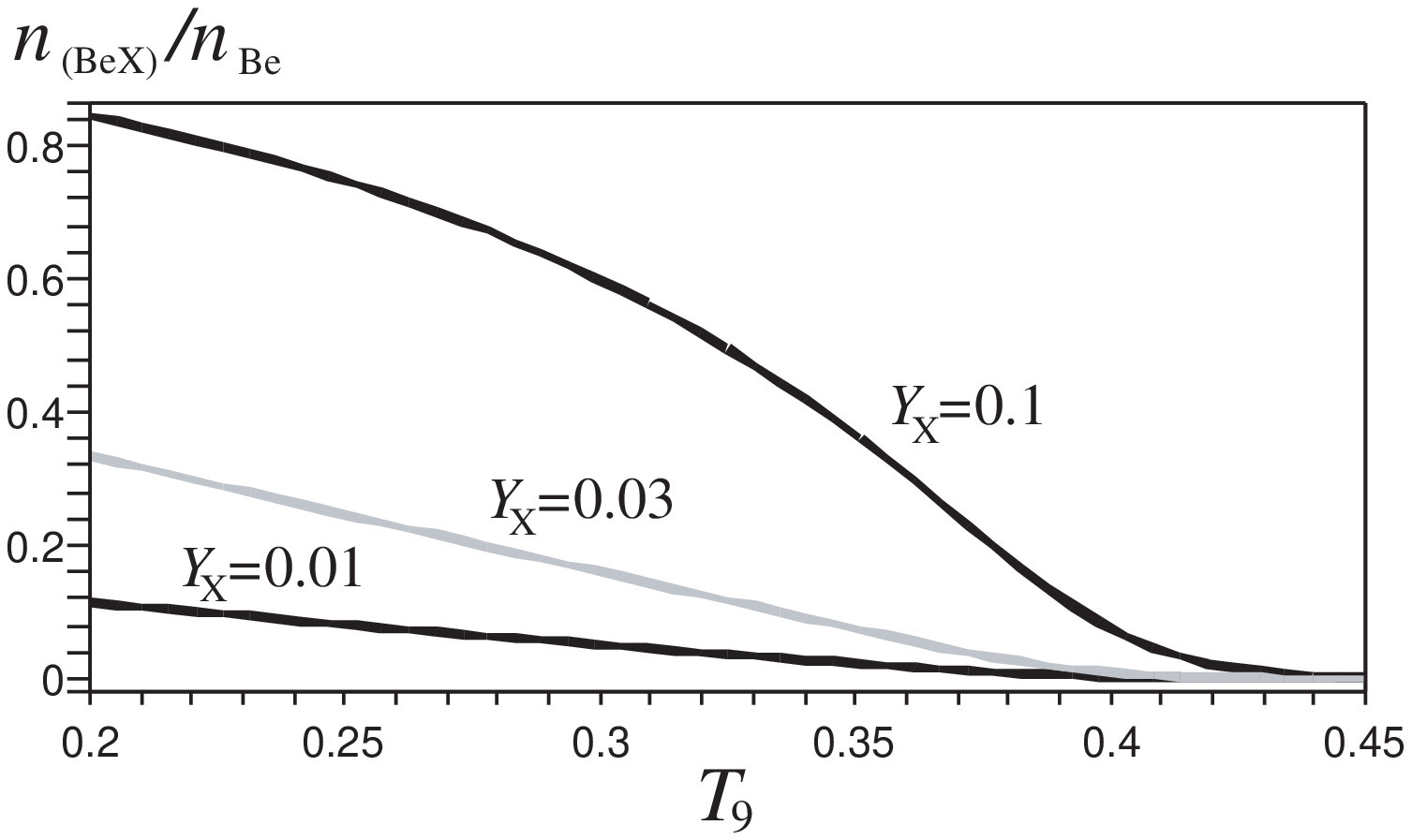}}
\centerline{\includegraphics[width=11 cm]{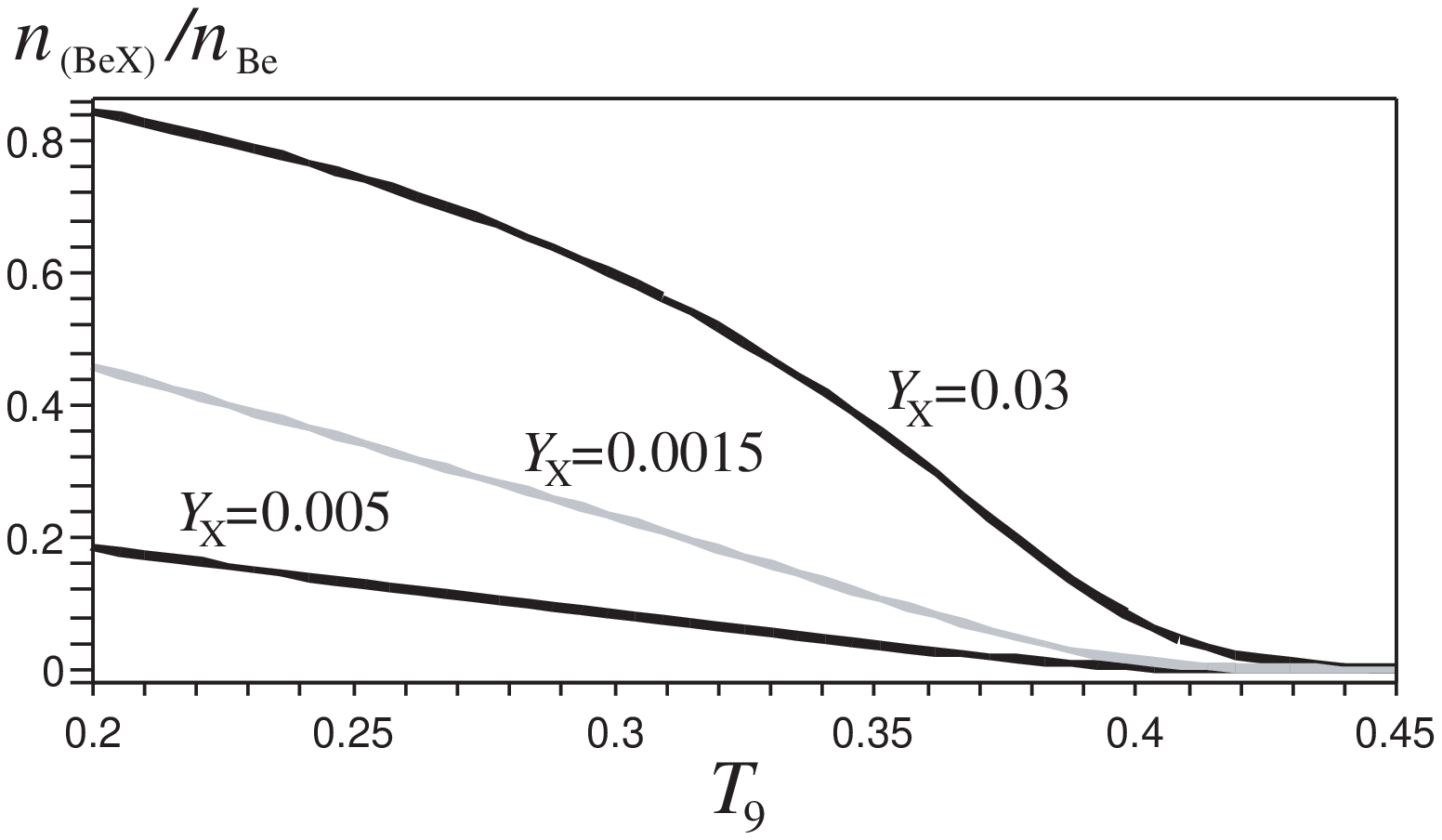}}
 \caption{\footnotesize Fraction of \bes\, locked in the bound state \bex\ as a function of temperature.
Top figure corresponds to a conservative choice of 
recombination cross section (\ref{sigma_tot}) and to 
three different values of $Y_X=n_X/n_B$: 0.1, 0.03 and 0.01.  Lower figure corresponds to recombination cross section 
enhanced by the $2s$ resonance (\ref{addon})  and $Y_X=0.03,~0.015,~0.005$. The long lifetime of 
\xm\ is assumed.  }
\label{f3} 
\end{figure}

Solutions for three different values of \xm\, abundance 
are plotted in Figure \ref{f3} for two choices of recombination rate, (\ref{sigma_tot}) and (\ref{addon}), 
with and without the contribution of the above threshold $2s$ resonance. It is important to keep in mind that 
these plots are for illustrational purposes only, 
as they neglect important nuclear effects that destroy \bex, which will be considered in subsequent sections. 
But even this simplified analysis shows that the abundance of $Y_X<0.005$ will be inconsequential 
for \bes\, abundance as less than 20\% of these nuclei form bound states at $T_9>0.2$. 
Comparison of two figures exemplifies the nuclear uncertainty, as the efficiency of the recombination 
rate may be enhanced by a factor of a few. 
Figure \ref{f3} also assumes constant abundance $Y_X$, {\em i.e.} the decays of \xm\, are neglected.
The generalization for a finite lifetime 
would result in a sharp drop in the number of \bex\ bound states at times much larger than  $\tau_X$.

\section{Catalyzed reaction channels} 

Once the bound state \bex\, is formed, several destruction channels get enhanced and 
new destruction channels open up. 

\subsection{ Catalysis of $p$-destruction.} The reaction \bes+$p\to\gamma+^8$B is one of the best studied 
in nuclear physics. Its rate is dominated by an $E1$ transition, and consequently depends on the 
third power of the frequency of the emitted photon. As a starting point for our estimates, we 
use the result of Ref. \cite{Baye} for 
the value the astrophysical $S$-factor at zero energy based on a two-particle model, 
\be
S_{\rm SBBN}(0) = \fr{5\pi}{9}\alpha\left(\fr{Z_1 A_2 - Z_2 A_1}{A}\right)^2\omega_{SBBN}^3 I^2(0),
\ee
where $I(0)$ is the radial nuclear integral, and $Z_i$, $A_i$ are charges and masses of nuclei
participating in $\gamma$-fusion. $^8{\rm B}$ produced in this reaction rapidly undergoes $\beta^+$-decay to \beet\, 
thus resulting only in two extra $\alpha$ particles. The  SBBN rate for this reaction is not sufficiently fast 
to reduce the amount of \bes, as it is about two orders of magnitude slower than the Hubble rate. 

The rate for this reaction would have been tremendously enhanced due to 
virtual photon exchange \cite{CBBN} if $(\besm X^-)+p\to{^8\rm B} + X^-$ were energetically allowed. In reality, this process 
can only go from $n\geq 4$ excited states of \bex\, and only a small fraction of \bex\, would be 
in these states. This does not mean, however, that the 
rate for the reaction involving the ground state is not enhanced. The process 
\be
(\besm X^-) + p \to (^8{\rm B}X^-) + \gamma,
\label{p-reaction}
\ee  
produces a photon which is $\sim 4.8$ times more energetic than in the non-catalyzed reaction, 
and has a reduced Gamow suppression.
 Assuming that nuclear radial integrals (stripped of their Coulomb suppression) 
have similar values in both the SBBN and CBBN cases, one arrives at the following, 
admittedly crude,  estimate for the enhancement of the 
catalyzed $S$-factor for the reaction in (\ref{p-reaction}):
\be
S_{\rm CBBN}(0) \sim  S_{\rm SBBN}(0)\left(\frac{\omega_{\rm CBBN}}{\omega_{\rm SBBN}}\right)^3
\frac{1}{0.37^2}\simeq 700\times S_{\rm SBBN}(0) \sim 15 {\rm keV~bn},
\ee
where  for the SBBN $S$-factor we use $S_{\rm SBBN}(0)=21$ eV bn, and 0.37 accounts for 
the change in $(Z_1 A_2 - Z_2 A_1)/A$ factor.

It turns out that in addition to the non-resonant part of the cross section, 
there are more important contributions from resonances. 
The most important resonant process is given by the following transition, 
\be
(\besm X^-) + p \to (^8{\rm B}X^-,~ n=2,l=1) \to (^8{\rm B}X^-)+ \gamma;~~~ E_R = 167~{\rm keV}
\label{Bepres},
\ee
where the resonant energy is given relative to the \bex+$p$ continuum and the 
input value of 2.64 fm is used for the charge radius of the $^8$B nucleus. 
The $E1$ electromagnetic width of the $(^8{\rm B}X^-,~ n=2,l=1)$ resonance 
can be easily calculated to be equal to approximately 10 eV. 
Using this information, and assuming the narrow resonance, we can predict the thermal 
rate for this process as 
\be
\langle \sigma_p v\rangle \simeq  1.6\times 10^8 ~ T_9^{-2/3}\exp(-8.86/T_9^{1/3})
+1.6\times 10^6~ T_9^{-3/2}\exp(-1.94/T_9).
\label{protons}
\ee
This is a very large rate for a reaction involving $\gamma$ in the final state.
The resonant part of this reaction dominates, and the $p$-burning of 
\bex\ remains faster than the Hubble rate until $T_9= 0.2$.

\begin{figure}[htbp]
\centerline{\includegraphics[width=11cm]{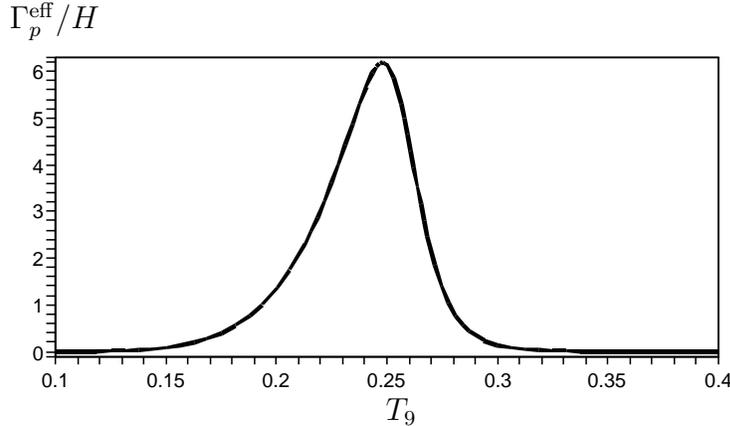}}
 \caption{\footnotesize Effective rate for proton destruction of (\bes\xm), that becomes efficient if the 
ratio $\Gamma_{p}^{\rm eff}/H$ is on the order or larger than unity. Even though the rate for 
$(\besm X^-) + p \to (^8{\rm B}X^-) + \gamma$ is much faster than Hubble rate, the reverse rate makes 
this destruction mechanism inefficient for $T>30$keV. 
}

\vspace{-7cm} \hspace{3cm} $\Gamma_{p}^{\rm eff}/H$

\vspace{4.7cm} \hspace{8cm} $T_9$

\vspace{1.5cm}
\label{effective} 
\end{figure}

Once the (\bes\xm) has being converted to $(^8{\rm B}X^-)$, the two important processes may occur.
One is the reverse rate $(^8{\rm B}X^-)+\gamma\to (\besm X^-) + p$, that brings back \bes, and the 
other process is the $\beta$-decay of $(^8{\rm B}X^-)$, for which we will use 
the standard lifetime value of $^8{\rm B}$, $\tau_{\rm B} = 1.11$ sec. At high temperatures,
the reverse rate reduces the efficiency of $p$-destruction very significantly, as pointed out by K. Jedamzik
in arXiv:0707.2070. 
Since the destruction of $(^8{\rm B}X^-)$ is very rapid compared to the Hubble scale, either by the 
beta decay or by the reverse reaction $\Gamma_\gamma$, one can employ the dinamical equilibrium 
assumption for the concentration of $(^8{\rm B}X^-)$, which leads to the following effective 
destruction rate of (\bes\xm) by the proton reactions,
\be
\Gamma_{p}^{\rm eff}(T) = \fr{\tau_{\rm B}^{-1} \langle \sigma_p v\rangle n_p}{\tau_{\rm B}^{-1}+\Gamma_\gamma}.
\label{gamma_eff}
\ee
Calculating $\Gamma_\gamma$ according to the standard procedure \cite{reactions}, 
we plot the effective destruction rate in units of Hubble rate in Figure \ref{effective}.

Figure \ref{effective} shows that in interval of temperatures from 20 to 30 keV the 
proton burning of (\bes\xm) is very efficient. 
After $\beta$-decay
most of $(^8{\rm B}X^-)$ bound states would transform to (\beet\xm), which would remain 
stable until the decay of \xm, which in turn will be immediately followed by the decay \beet$\to$2\hef. 
As an interesting side remark, we observe that intermediate (\beet\xm) bound states  
could potentially lead to a new primordial source of $^9$Be via the catalyzed 
$(^8{\rm Be}X^-)+n\to {\rm ^9Be} +X^-$
reaction. 

\subsection{ Catalysis of $n$-destruction}

It is well-known that the reaction $\besm+n\to \lisvm+p$ is the main mechanism for depleting \bes\ abundance in 
SBBN. This is a reaction with one of the largest cross sections known to nuclear physics with $A\leq8$. 
It is dominated by a wide near-threshold resonance in \beet\ ($E_{\beetm}=18.9$MeV) 
with the neutron and proton widths being 0.22 and 1.41
MeV respectively \cite{AD}. With such large widths, the binding effects to \xm\ are not going to 
change the rate for the $(\besm X^-)+n\to (\lisvm X^-)+p$ reaction in any substantial way. 

The enhancement of the $\besm+n\to 2$\hef\ was mentioned in \cite{french} as one of the 
important destruction mechanisms if its rate is scaled up by an arbitrary enhancement factor of a 
hundred. Ref. \cite{french} concludes that such rescaling is unlikely in this channel. One of the 
papers that investigated the effects of the bound states with charged particles \cite{KT} argues that 
the rate for the $(\besm X^-)+n\to  X^-+2$\hef\ process,
\be
2\times 10^4(1+3.7~ 10^3T_9),
\label{Benaa}
\ee 
could indeed be greatly enhanced by up to three 
orders of magnitude because \bes\ on the orbit of the \bex\ bound state has large kinetic energy that effectively 
replaces $T_9$ in the formula for the rate. 
The absence of $T_9$ in the rate above 
would necessarily mean $s$-wave rather than $p$-wave scattering, which can only happen if $E_R=18.9$ MeV
resonance
 dominates the cross section. The reason why the $s$-wave part  is relatively small 
in the SBBN rate (\ref{Benaa}) is because the symmetry of the final state forbids the 
photonless decay to 2\hef\ and allows only $\besm+n\to \beetm^*\to\gamma+2$\hef. It is clear that in the 
CBBN process, internal conversion may occur, enabling a photonless 
$s$-wave reaction $(\besm X^-)+n\to  X^-+2$\hef. The overall enhancement of the $s$-wave ($T_9$-independent) part 
of (\ref{Benaa}) is expected to be on the order
of $(\lambda_{\rm real}/\lambda_{\rm virtual})^3$, which for this transition is not likely to exceed 
$10^3$. Based on this argument, we conclude that the rate for $(\besm X^-)+n\to  X^-+2$\hef\ is 
catalyzed rather moderately, and would not exceed $\sim$ few$\times 10^8$. Such a rate is still subdominant to 
$\besm+n\to \lisvm+p$, and the destruction of \bex\ by neutrons is less efficient than by protons (\ref{protons}),
especially taking into account that neutrons are less abundant than protons by approximately six orders of magnitude at 
relevant temperatures.

\subsection{ Break-up of \bes\ in the decay of \xm.} 
New channels of \bes\, break-up 
come from the decay of the \xm\, particle inside the \bex\ bound states. There are three different possibilities:\\
A: \bex$\to$\bes\, + {\rm products of decay + hard} $\gamma$ $\to$ \het+\hef+...\\
B: \bex$\to$\bes($ E\gg T$)\, + {\rm background particles}  + ... $\to$ \het+\hef+...\\
C: (\bes$^*$\xm)$\to$ \het+\hef+...\\
In case A, the process of decay of \xm\, and/or the charged products in the decay chain produce 
a photon with energy in excess of the photodisintegration threshold for \bes, $E_\gamma > E_{thr}=1.59$ MeV, 
that hits the \bes\ nucleus and destroys it. 
This is a hard photon with respect to the thermal bath, $T\sim O(30 \; {\rm keV})$ , but is considered a soft photon 
with respect to the energies of the \xm\, decay in type I models, where the typical energy release is $O(100 \; 
{\rm GeV})$. 
The process B is much simpler. The decay of \xm\, occurs instantaneously, so that the emergent \bes\ nucleus 
has a distribution over the range of momenta given by the Fourier transform of the wave function 
from Figure \ref{f1}. Since the typical kinetic energies of \bes\ on orbit inside \bex\ are on the order of 
a few MeV, the break-up of the recoiling \bes\ by other nuclei in the primordial plasma  
becomes more efficient as the Gamow suppression is effectively lowered. The final option C is due 
to the polarization 
of \bes\, by the electric field exerted by \xm\, inside \bex. Effectively, the \bes\ within the bound state is 
represented by the ground state of {\em isolated }
\bes\, admixed with a combination of all the excited states, most of which (with the exception of $\besm_{1/2}$) belong to the 
continuum. Upon the instantaneous decay of \xm\, these extra ``polarized" pieces of the \bes\, wave function 
lead to a decay to the \het+\hef\, continuum. Mechanism C exists for both types of models, with or without large 
energy injection.

To estimate the efficiency of channel A, we take a model of type I with the assumption that the decay produces a single 
charged track with energy $E_{max}\gg 1.59$ MeV. We estimate the number of Weizs\"acker-Williams  photons $dn_\gamma(E)$ 
with $ E\ll E_{max}$, determine the effective flux for the \bes\, target and obtain the following probability 
for the radiative break-up,
\be
P_{rad ~ br} \simeq \int_0^{\infty} dr |\psi(r)|^2 
\int_{E_{thr}}^{E_{max}} \sigma_\gamma(E) dn_\gamma(E)
\label{breakup}
\ee
Numerical evaluation of (\ref{breakup}) with the input of cross section for $\besm+\gamma\to$\hef+\het\
produces the following estimate,
\be
P_{rad ~ br} \sim 10^{-4} ~~~~{\rm for }~~~E_{max}=100~{\rm GeV},
\ee
which forces us to conclude that the radiative breakup of the recoiling \bes\, is rather inefficient. 

Estimates of process B depend very sensitively on the rate of slow down of \bes. For the relevant 
temperature range (on the order of 30keV), we determine that it takes 
approximately $10^{-6}$sec for an MeV-energy 
\bes\, to thermalize. This time interval is not sufficient for a proton  induced break-up
to happen, \bes+$p\to^9$B+$\gamma$, even though the rate for the latter is enhanced 
relative to thermal rate due to a larger center of mass energy of $\besm+p$ system. 
The estimated probability for the break up in mechanism 
B during the thermalization time remains negligibly small, and we conclude that process B is also inefficient. 

Finally, in option C we need to estimate the admixture of excited states to the ground state 
of \bes\ nucleus by the electric field of \xm\ particle. Schematically, this probability is  
\be
\sum\left|\frac{({\bf d E})_{0i}}{E_0-E_i}\right|^2,
\ee
and the sum runs over all continuum states of \bes. Taking a typical value of $({\bf d E})_{0i}\sim 1$ MeV, 
and energy denominator comparable to Gamow energy, $E_i-E_0\ge 10$ MeV, we estimate that the total probability of 
admixture of excited states is only about 1\%.  Thus, we are forced to conclude that 
all mechanisms related to the decay of \xm\ within the 
bound state \bex\ do not lead to a significant depletion of \bes\ abundance. 

\subsection{ Internal conversion: $\besm+X^-\to \lisvm+X^0$.} 
\begin{figure}[htbp]
\centerline{\includegraphics[width=6 cm]{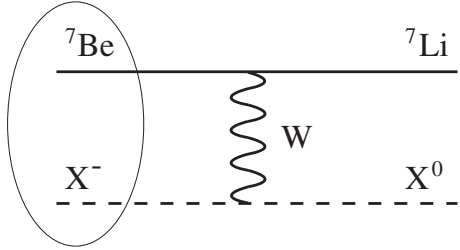}}
 \caption{\footnotesize Internal conversion of beryllium into lithium in 
type II models due  to $W$-exchange. \lisv, being intrinsically more fragile than 
\bes\,, is subsequently destroyed.}
\label{f4} 
\end{figure}

Any process that converts \bes\, to \lisv\, early enough would lead to a 
suppression of the total lithium abundance, \lisv+\bes, as \lisv\, is significantly more 
fragile. 
In type II models of charged relics, where their longevity is ensured by the 
small energy splitting between \xm\, and \xz, it is natural to expect the existence of 
weak charged current between \xz\, and \xm,~\xp. In that case the process shown in Figure \ref{f4}
becomes possible,
\be
\bexm\to \lisvm\,+X^0.
\label{conversion}
\ee
This process is analogous to a usual electron capture  process, 
(\bes ~$e^-)\to \lisvm+\nu$. The main difference is that the rate for 
(\ref{conversion}) is many orders of magnitude larger than the standard capture
of $e^-$ inside the beryllium atom. 
There are two primary reasons for that. The usual capture rate is proportional 
to $|\psi(0)|^2\sim (Z_{\rm Be}\alpha m_e)^3$, {\em i.e.} to 
the probability density of finding an electron at the 
same point as \bes. For the reaction (\ref{conversion}) this factor should be taken to be 
$\sim r_{\rm Be}^{-3}$, since the average distance between \bes\, and \xm\, is less 
than the nuclear radius $r_{\rm Be}$. The second source of enhancement is in the final state 
momentum of particles involved in the transition. More specifically, due to different kinematics, 
the square of the neutrino momentum $p_{(\nu)}^2$ must be substituted for  $m_{\rm Li} p_{\rm Li}$.
Altogether this amounts to an extremely large enhancement factor,
\be
N_{enh} \sim (r_{\rm Be}Z_{\rm Be}\alpha m_e)^3 \times (m_{\rm Li} p_{\rm Li}/p_{(\nu)}^2)\sim 10^{15};~~~
\Gamma_{int~conv} \sim 10^{-6}~{\rm eV}.
\label{fast}
\ee
This must be compared to the rate of the normal capture, which is in the range 
of $O(10^{-6}$sec$^{-1}$)$\sim O(10^{-21}$eV). 

Comparing rate (\ref{fast}) with the photodissociation rate of \bex\, we conclude that 
(\ref{fast}) becomes dominant for $T_9 \simlt 1$, which covers the whole range 
of temperatures relevant for the lithium abundance problem. Because of this, in type II models  
the rate of recombination determines the \bes$\to$\lisv\, conversion rate induced by \xm, and the 
actual width (\ref{fast}) does not enter Boltzmann equations, as it is {\em larger} than the 
rest of the reaction rates.

\subsection{ Energy injection through the annihilation of \xm and \xp.} The impact of annihilating 
particles on the BBN predictions was recently considered in detail in Ref. \cite{annih}. 
In the CBBN scenario, the injection of energy due to the annihilation of \xm\, and \xp\, can be 
calculated with reasonable accuracy. At temperatures much below the binding energy of the (\xm\xp) system,
this annihilation proceeds via the formation of a positronium-like bound state. 
This recombination process is followed by the rapid annihilation of \xm\, and \xp\, within the bound state. 
The value of binding energy 
$E_b = m_X\alpha^2/4$ is larger than 1 MeV for $m_X>100$ GeV. Therefore, at $T<100$ keV it is safe to ignore
the reverse process of photodisintegration of (\xm\xp). 

A straightforward calculation gives the amount of energy released per  \xm\ particle per time,
\be
\Gamma_{ann} =\langle\sigma_{(X^-X^+)}v_{rel}\rangle n_{X^+} 
\label{rate_ann}
\ee
Unlike the case of \bes+\xm\, recombination, the hydrogen-like (or more precisely, the positronium-like)
expressions for the cross sections are appropriate in this case,
\be
\langle\sigma_{(X^-X^+)}v_{rel}\rangle = 
\fr{2^9\pi^2}{3\exp(4)}\fr{\alpha}{\alpha^2\mu^2}\fr{2E_b^2}{\mu^2}\left\langle \fr{1}{v_{rel}}\right\rangle
=
\fr{2^{10}\pi^{3/2}\alpha^3}{3\exp(4)m_X^{3/2}T^{1/2}},
\label{xmxp}
\ee
where $\mu$ is the reduced mass, $\mu=m_X/2$, and $E_b = \alpha^2m_X/4$. Notice that it is the mass of the $X$-particle, rather than the 
nuclear mass, that sets the scale of recombination cross section, and consequently the rate (\ref{rate_ann}) 
is subdominant to the Hubble rate. Therefore the formation of (\xp\xm) positronium-like state 
with its subsequent annihilation does not lead to an 
appreciable depletion of $n_X$. It is instructive to calculate the actual size of this rate at 
temperatures relevant for the lithium problem:
\be
\langle\sigma_{(X^-X^+)}v_{rel}\rangle_{(T_9=0.3)} 
\simeq 100~{\rm pbn}\times \left(\fr{500~{\rm GeV}}{m_X}\right)^{3/2}.
\ee
This is to be compared with the sub-picobarn size of cross section for the annihilation of 
neutralino particles. Therefore the effect from (\xp\xm)-induced energy injection is expected to be 
much more important for the BBN than neutralino annihilation \cite{annih}. 

\par
It is customary to quantify the effects of the unstable particles on BBN by the energy released
per photon $\xi$ normalized on 1 GeV. A similar quantity for annihilating particles would 
read as
\be
\xi = \frac{2m_X}{1{\rm GeV}}Y_X^2\eta_B^2\times \int_{T_1}^{T_2}
\fr{\langle\sigma_{(X^-X^+)}v_{rel}\rangle  n_\gamma}{TH(T)}dT,
\label{xi}
\ee
where $T_1$ to $T_2$ is the relevant temperature interval, which for our estimates can be 
taken to be from $\sim$ 10 to 40 keV. Above these temperatures, the impact of energy injection will be washed away
by thermal nuclear reactions. 

Substituting (\ref{xmxp}) into (\ref{xi}), and normalizing $Y_X$ and $m_X$ on their typical 
``electroweak scale-inspired" values, we arrive at 
\be
\xi = 2.2 \times 10^{-12} \left(\fr{500~{\rm GeV}}{m_X}\right)^{1/2} \left(\fr{Y_X}{0.02}\right)^2.
\label{release}
\ee
This small energy release per photon is inconsequential for the BBN predictions if the energy is released via 
the electromagnetic radiation, {\em e.g.} \xm+\xp$\to$(\xm\xp)$\to \gamma\gamma$. However the hadronic energy 
release via \xm+\xp$\to$(\xm\xp) $\to{q}\bar{q}$ even at this $O(10^{-12})$ level is capable of 
suppressing \lisv. For example, provided that the hadronic branching ratio is order 1, the estimate (\ref{release})
is consistent with a factor of a few decrease in \lisv\, abundance reported in \cite{jj}.
It is important to note that the realistic scaling of (\ref{release}) with mass is $m_X^{3/2}$, as 
the standards arguments for the annihilation at the freeze-out force $Y_X$ to scale linearly with $
m_X$ when it becomes heavier than the electroweak scale. Thus for heavier 
$m_X$ the effect is even more pronounced. 
\par
Subsequent decays of \xm and \xp\, may or may not lead to a larger energy 
release than residual annihilation. 
In models of type I, the decay would typically provide larger effect, while in models 
of type II the annihilation is more important. In any event, this subsection 
shows that even for type II models, in which the decays of 
\xm\, and \xp\, do not release a significant amount of energy, the non-thermal component of BBN is  still 
unavoidable if the hadronic branching in the annihilation of (\xm\xp) is appreciably large.

\section{Evolution of \bes\, + \lisv\, at $T\simeq 30$ KeV}

The standard mechanism of generating \bes+\lisv\, at the WMAP-suggested value for $\eta_b$ is relatively simple. 
The main reactions that determine total lithium abundance are
\be {\rm SBBN}: ~~~~
\hetm+\hefm\ \to \besm +\gamma; ~
\besm + n \to \lisvm + p; ~ \lisvm+p\to 2\hefm {\rm ~or~ D} +\lisxm.
\label{SBBNreac}
\ee
The first reaction generates \bes\, while the combination of the second and third burn it. 
The secondary creation mechanism is due to \hef+$^3$H$\to$\lisv+$\gamma$, while the secondary destruction 
mechanisms are given by $\besm+p\to {\rm ^8B} +\gamma$ and $\besm+{\rm D}\to p+2\hefm$.
All of these processes were discussed at length in the existing BBN literature \cite{french,Rich}. 

The CBBN creates several additional destruction mechanisms that were described in the previous 
sections. Below we list main CBBN processes:
\be
~~~~~~~~~~{\rm CBBN}: ~~~~ \besm+X^-\leftrightarrow (\besm X^-)+\gamma;~~ \lisvm+X^-\leftrightarrow (\lisvm X^-)+\gamma;
\ee
$$
{\rm Type ~ I~ and ~ II}: ~~~~ (\besm X^-)+p\leftrightarrow ({\rm ^8B}X^-) +\gamma; ~~~
({\rm ^8B}X^-)\to ({\rm ^8Be}X^-).$$
$$
(\besm X^-)+n\to (\lisvm X^-)+p;
 ~~(\lisvm X^-)+p\to X^-+2\hefm  ~~{\rm or }~~  X^-+{\rm D}+ \lisxm. $$ 
$$
{\rm Type ~  II ~ only }:~~~~ (\besm X^-) \to \lisvm+X^0; ~~({\rm ^8B} X^-) \to \beetm+X^0.
$$
Besides these main channels, we also included the \bex\, destruction by D-burning. In $(\lisvm X^-)+p$ 
and \bex+D reactions the only change implemented relative to the SBBN rate 
was in the Coulomb penetration factor. 

The models of Type II have an additional advantage over type I because of the internal conversion process 
$(\besm X^-) \to \lisvm+X^0$. As soon as the rate for this process becomes dominant 
over the Hubble rate, one can consider the photorecombination of \bes\, and \xm\, as leading directly to 
\lisv,
\be
 \besm+X^-\to(\besm X^-)+\gamma\to \lisvm+X^0+\gamma,
 \ee
with the rate given by Eqs. (\ref{sigma_tot}) or (\ref{addon}).
\par
The actual calculations of the abundance are performed using only the Li-Be part of the 
BBN code. Since the abundances of these elements are extremely tiny, their "backreaction"
on \hef, D and \het\, is negligible. That is, the destruction or creation 
mechanisms of \bes\, and\lisv\ do not manifest themselves in any noticeable change in the lighter elemental 
abundances. Thus the CBBN effect on lithium and beryllium can be studied separately, and we do so 
by creating a corresponding code with the use of MAPLE 9.5. The network of reactions given in (\ref{SBBNreac}),
which we take from well-established sources \cite{reactions}, and the input of D$(T)$, $^3$He$(T)$, etc.  
from the full SBBN code gives an approximation to the freeze-out 
abundance of total lithium as 
\be
(\lisvm^{\rm tot})_{\rm SBBN}\equiv (\besm+\lisvm)_{\rm SBBN} = 4.1\times 10^{-10},
\ee
in excess of the Spite plateau value, and indeed in agreement with more elaborate treatments of 
Refs. \cite{french,Keith,Rich}.

\begin{figure}[htbp]
\centerline{\includegraphics[width=14cm]{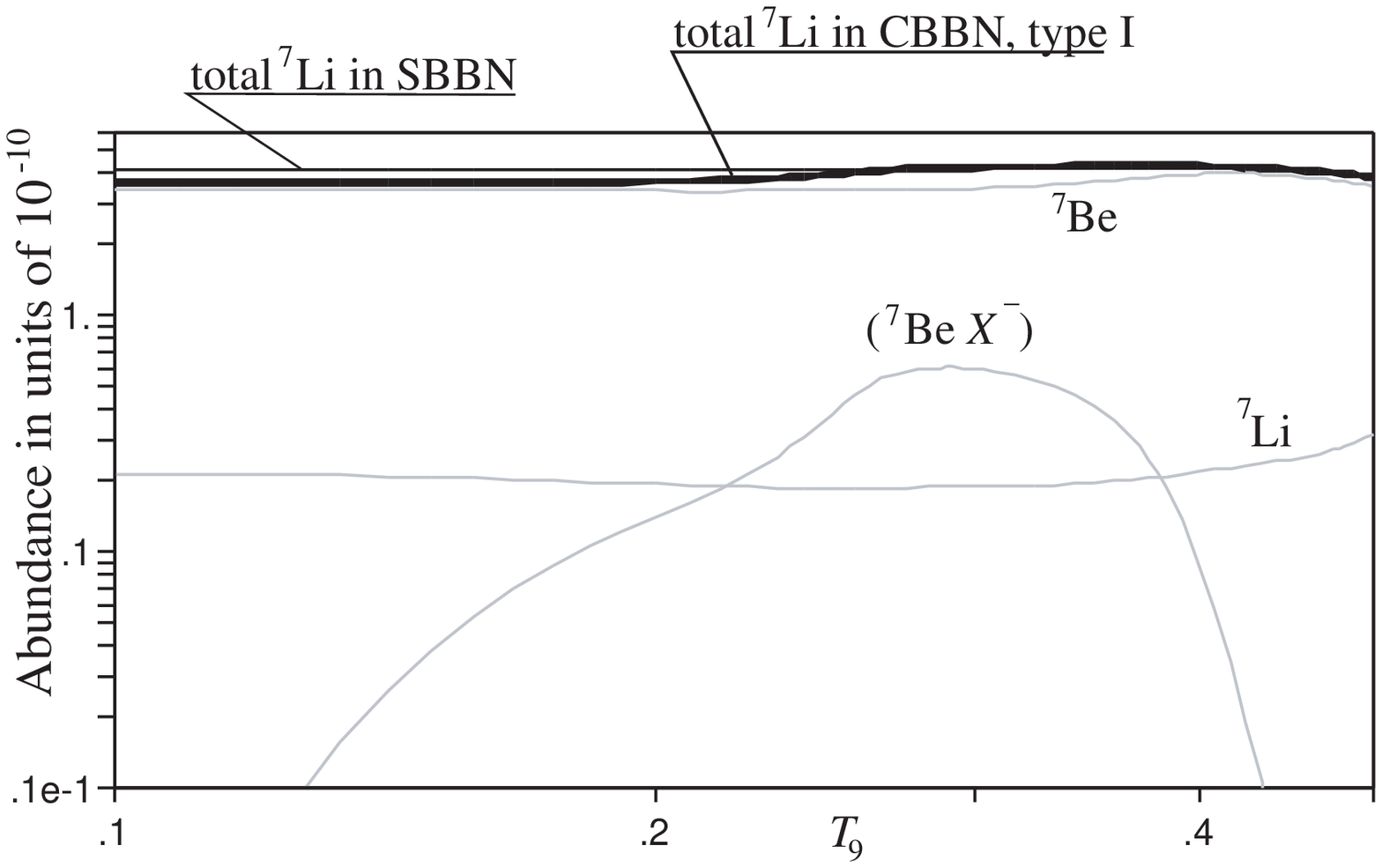}}
\centerline{\includegraphics[width=14cm]{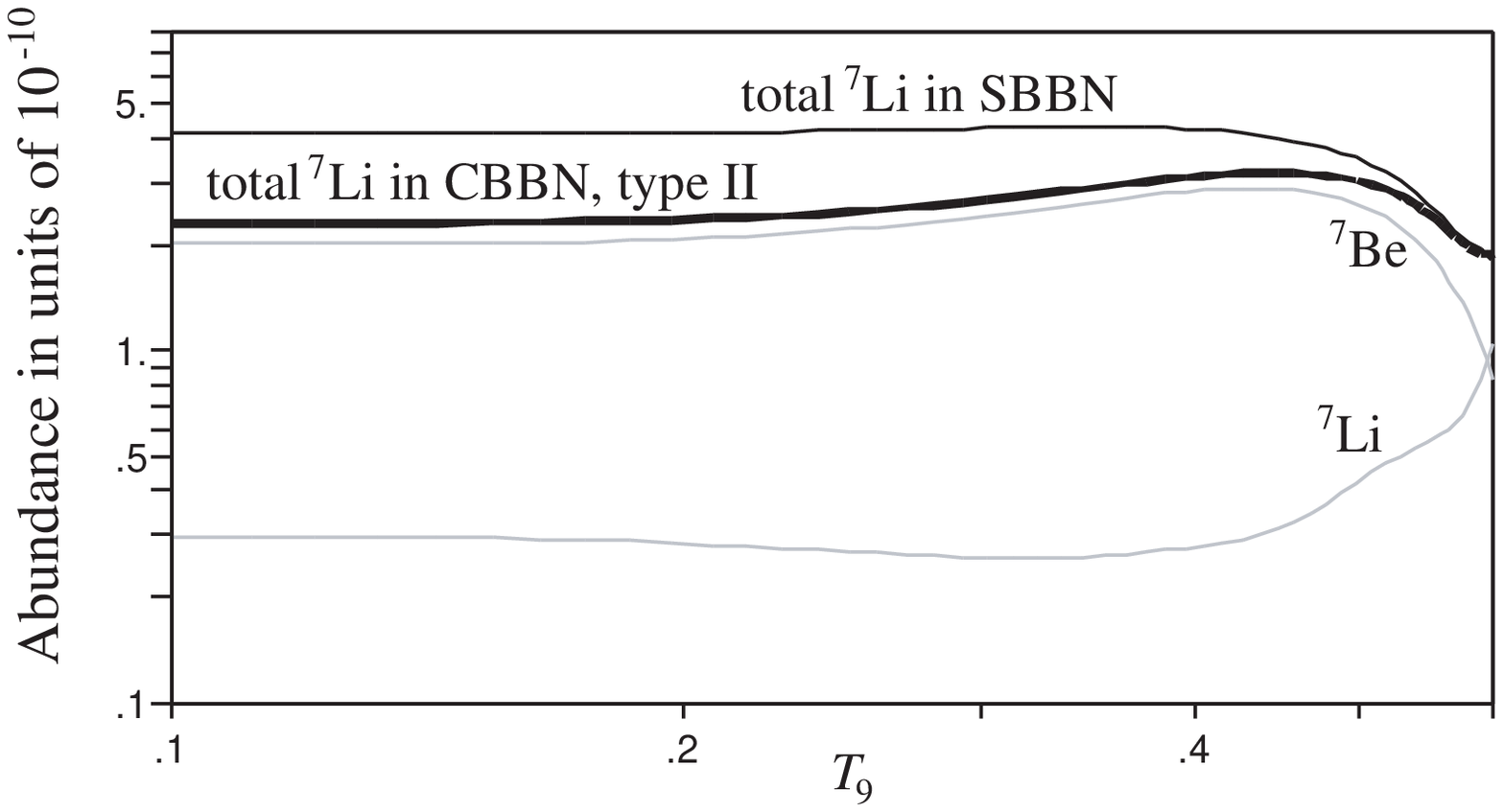}}

 \caption{\footnotesize The abundance of lithium and beryllium in two types of the CBBN 
models for the sample values of the CBBN input parameters, $Y_X=0.05$ and $\tau_X =2000$ sec,
and the choice of recombination cross section (\ref{sigma_tot}). 
The total lithium abundance is $3.7\times 10^{-10}$ for Type I and $2.3 \times 10^{-10}$ for Type II model. 
The individual abundances of \bes, \lisv, \bex\, and the SBBN curve for \lisv+\bes\, are also shown.
All abundances are given relative to hydrogen.}
\label{f5} 
\end{figure}

\begin{figure}[htbp]
\centerline{\includegraphics[width=14 cm]{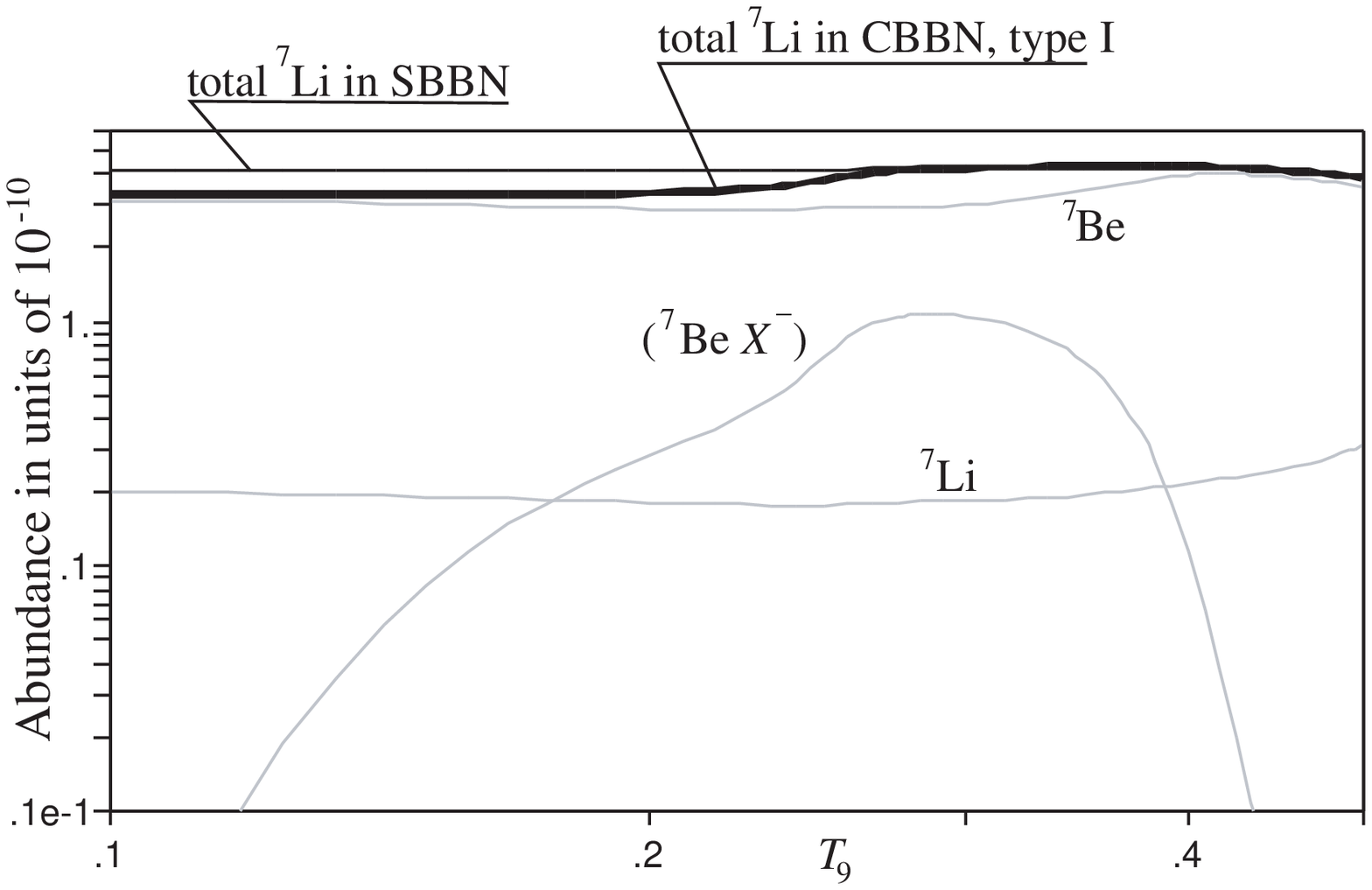}}
\centerline{\includegraphics[width=14 cm]{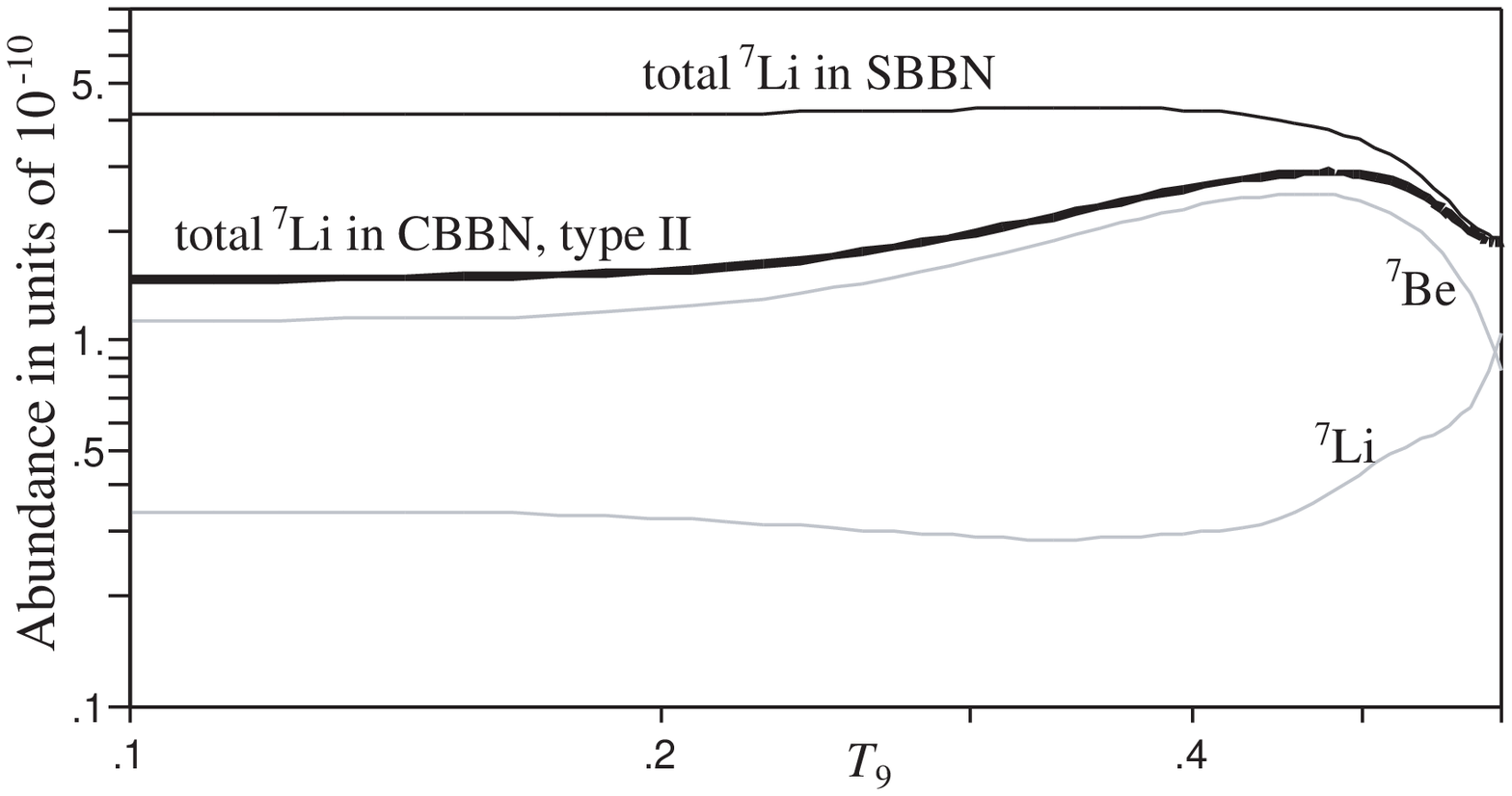}}

 \caption{\footnotesize The abundance of lithium and beryllium in two types of the CBBN 
models for the sample values of the CBBN input parameters, $Y_X=0.05$ and $\tau_X =2000$ sec,
and the choice of recombination cross section (\ref{addon}). 
The total lithium abundance is $3.3\times 10^{-10}$ for Type I and $1.5 \times 10^{-10}$ for Type II model. 
The individual abundances of \bes, \lisv, \bex\, and the SBBN curve for \lisv+\bes\, are also shown.
All abundances are given relative to hydrogen.}
\label{f5a} 
\end{figure}

We then extend our code by inclusion of the CBBN reaction chains, 
and do so separately for Type I and Type II models. In this paper we {\em do not} treat the 
non-thermal BBN processes due to the energy release associated with the decay and annihilation 
of \xm\ and \xp\ particles. Such processes indeed could be important in type I models, 
and lead to significant modifications of elemental abundances. They are inevitably more 
model dependent, and were previously considered elsewhere \cite{metastable,jj,Cy}. 
We also perform calculations for two forms of the recombination rate, with and without the 
effect from the intermediate $2s$ state in (\bes\xm). 

The temperature evolution of \bes, \lisv\ and their bound states with \xm\ are shown 
for typical 
values of the lifetime and abundance of \xm\, is presented in Figures \ref{f5} and \ref{f5a}. 
The total freeze-out abundance of lithium is changed in the following manner:
\begin{eqnarray}
\label{without}
{\rm Eq.}~(2.11);~~Y_X=0.05;~~ \tau_X = 2000{\rm s} ~~ \Longrightarrow~~ \lisvm^{\rm tot}_{\rm CBBN}=\left\{
\begin{array}{c}3.7\times 10^{-10}~~ {\rm for ~ type~ I} \\2.3 \times 10^{-10}~~ {\rm for ~ type~ II}
\end{array}
\right.
\\
\label{with}
{\rm Eq.}~(2.13);~~Y_X=0.05;~~ \tau_X = 2000{\rm s} ~~ \Longrightarrow~~ \lisvm^{\rm tot}_{\rm CBBN}=\left\{
\begin{array}{c}3.3\times 10^{-10}~~ {\rm for ~ type~ I} \\1.5 \times 10^{-10}~~ {\rm for ~ type~ II}
\end{array}
\right.
\end{eqnarray}
where (\ref{without}) corresponds to a conservative choice of recombination cross section (\ref{sigma_tot}), 
while (\ref{with}) takes into account the contribution of $2s$ resonance (\ref{addon}) in the assumption 
that this resonance is just above the energy threshold. 
The overall reduction of lithium abundance for this choice of parameters is 10\% in type I model 
and almost 50\% for the type II for the conservative value of recombination cross section, and 25\% 
and a factor of $\sim3$ for the recombination cross section enhanced by the $2s$ resonance.

The behavior of individual curves in Figures \ref{f5} and \ref{f5a}  can be easily explained. 
The SBBN curve remains almost flat below $T_9=0.4$, while the CBBN introduces an extra suppression, resulting 
in a slight decrease of the total lithium abundance below $T_9=0.4$. It is important to note that the overall 
abundance of \bex\ remains very low in models of type II, being controled by the fast
internal capture process. In type I it can be significant at $0.25\la T_9\la 0.4$ but 
then the $p$-burning of (\bes\xm) makes a noticeable "dent" in the 
(\bes\xm) abundance around $T_9\sim 0.25$.  At later time, all 
$X$-containing nuclei decline exponentially due to the decay of 
\bex\ at $T_9<0.2$. At the same time, \lisv\ abundance increases slightly below $T_9=0.3$ 
because the $p$-burning of \lisv\ becomes less efficient. 

\begin{figure}[htbp]
\centerline{\includegraphics[width=14 cm]{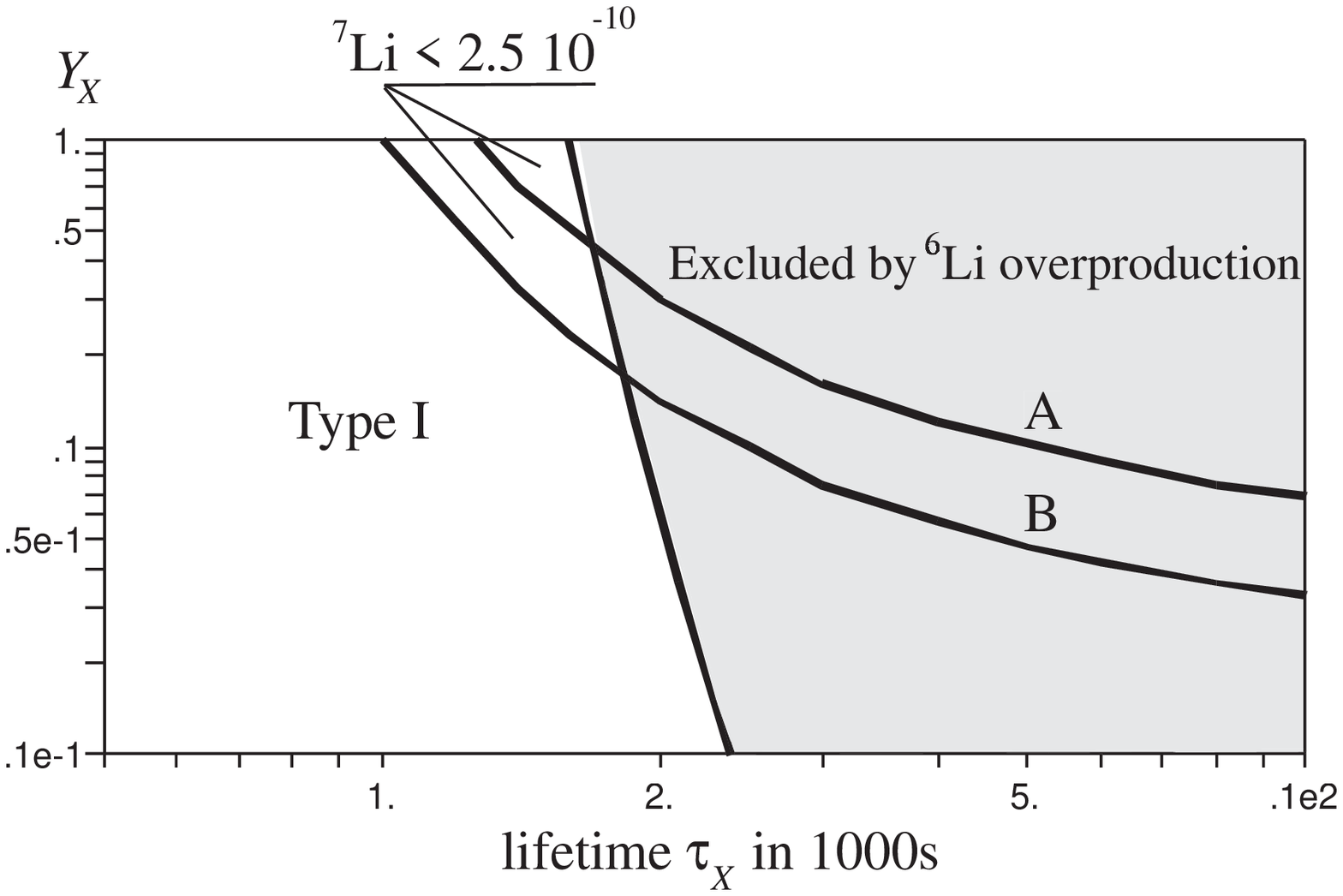}}
\centerline{\includegraphics[width=14 cm]{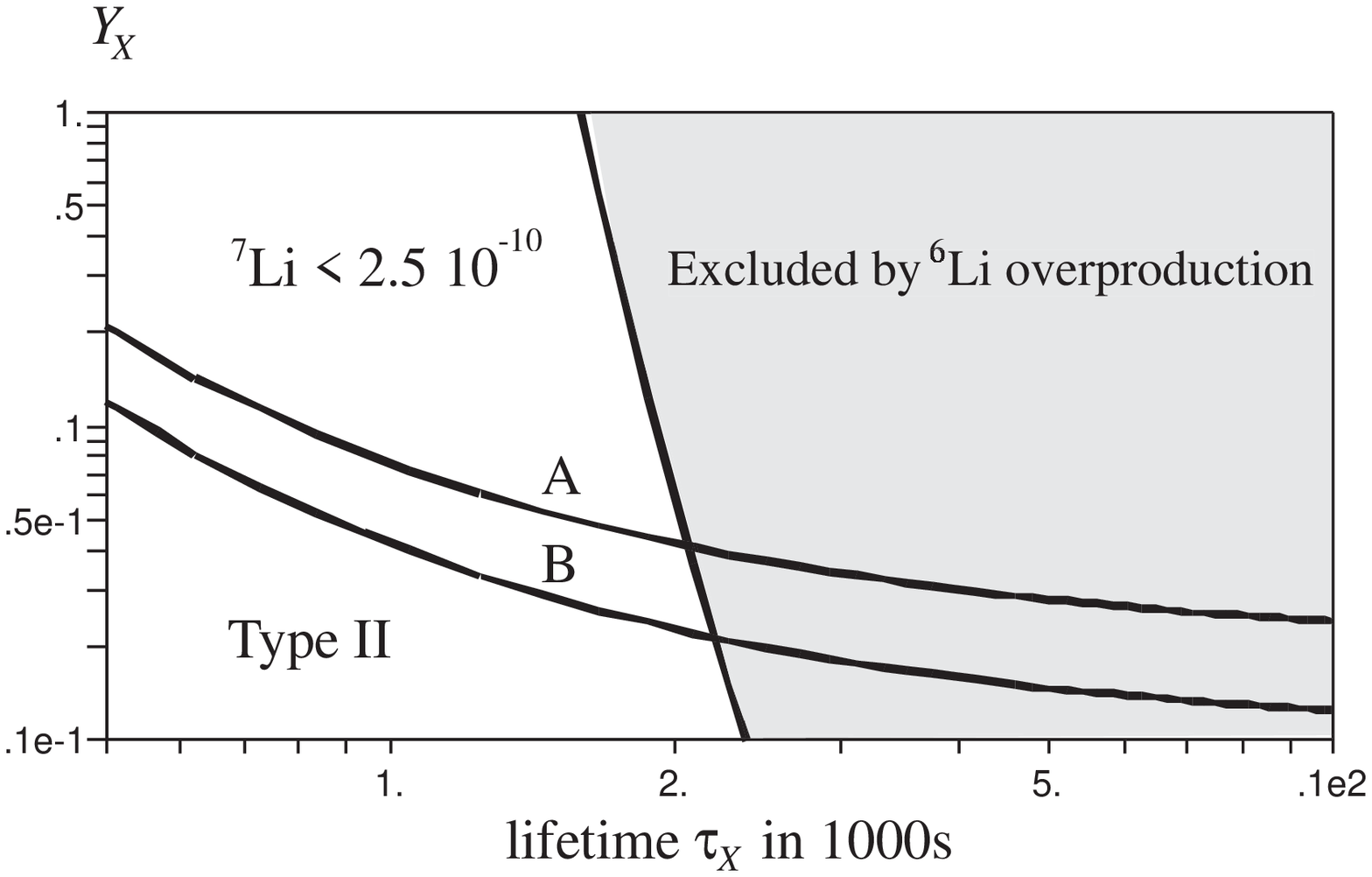}}

 \caption{\footnotesize $\{Y_X,\tau_X\}$ parameter space with CBBN Lithium abundance. 
The top panel corresponds to type I models, the lower panel corresponds to type II. 
Curves A and B are calculated with the use of (\ref{sigma_tot}) and (\ref{addon}) 
correspondingly. 
The \bes+\lisv\ 
overabundance problem is solved above curves A and B, with $(\lisvm^{\rm tot})_{\rm CBBN} < 2.5\times 10^{-10}$. 
The domain to the right from the  almost vertical line 
is excluded from the \lisx\ overproduction 
at $T_9 = 0.09$.  }
\label{f6} 
\end{figure}

Next we explore the parameter space of possible $\{Y_X,\tau_X\}$ values that would suppress 
total lithium abundance to an ``acceptable" level. It is clearly a rather contestable issue of 
what to call an ``acceptable level". We take this value to be $2.5\times 10^{-10}$ leaving some 
room for further possible reduction of lithium abundance in stars. The results of the scan through the 
parameter space is given in Figure \ref{f6}. 
This figure presents to models, with and without internal conversion, separately. 
One can see that the $Y_X\sim O(.05)$ and 
$\tau_X\sim (1-10)\times 10^3$s can provide an appreciable reduction of total lithium abundance. However the 
catalytic production of \lisx\ at $T_9\sim 0.09$ reduces the available parameter space quite 
significantly. Approximating the \lisx\ bound \cite{CBBN} by  

\be
Y_X(T_9=0.09)< 10^{-6}~ \Longrightarrow ~ Y_X\times\exp(-2.2\times 10^4{\rm s}/\tau_X)<10^{-6},
\ee
we find that only rather short-lived \xm\ particles,
\be
{\rm few}\times 100~{\rm s}~\la\tau_X\leq 2000~{\rm s}
\ee
can provide a reduction of \bes+\lisv\ abundance without producing too much \lisx\ via the 
formation of (\hef\xm) bound states. Overall, once can see that the inclusion of internal conversion of 
(\bes\xm) to \lisv\ and $X^0$ increases the efficiency of CBBN by a lot, allowing for a much more 
parameter space where both lithium problems are resolved by the catalytic means. The figures also show 
that the inclusion of the $2s$ resonance can increase the efficiency of the CBBN by a lot, 
and therefore a dedicated calculation of the nuclear levels in (\bes\xm) system with better than 
$\sim 50 $ keV accuracy is certainly warranted. For models of type I one can see the 
dramatic decrease in the efficiency of CBBN reduction of \lisv\ abundance for small values of  
$\tau_X$, while in models of type II this effect is less pronounced. It can be directly attributed to the
fact that $p$-buring of (\bes\xm) is only efficient for $T_9 \sim 0.25$, and most of the \xm\ with lifetimes
of the few hundred seconds would decay by then.

\section{ Discussion and conclusions}

Thermal catalysis of nuclear reactions by metastable heavy charged particles is a new 
way in which particle physics can affect primordial abundances. In the previous paper \cite{CBBN}
a new strong source of \lisx\ was pointed out, originating from the recombination of 
\hef\ nuclei and \xm\ at temperatures of about 8 keV. The bound on \lisx\ is so strong \footnote{Recent 
three-body calculation \cite{newCBBN} revises the original estimate of the catalytic enhancement of the 
$S$-factor from about $1.6\times 10^7$ to $2.1\times 10^6$, which does not alter main conclusions of 
Ref. \cite{CBBN} in any substantial way.  }
that 
it requires all \xm\ particles decay before that, $\tau_X<5000$ s. 
In this paper we analyzed earlier stage of nucleosynthesis at about 30 keV temperatures. 
During that period, the only stable bound state that can be formed is \bex. Its formation 
{\em reduces} the total \bes+\lisv\ abundance because this state is more fragile than \bes.
It is therefore tempting to speculate that the CBBN may provide a satisfactory solution 
to the lithium overproduction problem using the catalysis of thermal nuclear reactions. 

\par
We find that indeed a CBBN reduction of the total lithium abundance is 
possible for some values of the abundance-lifetime $\{Y_X,\tau_X\}$ parameter space. This 
reduction happens more efficiently in models with nearly degenerate \xm,\xp\ and $X^0$ states,
connected by the weak current, enabling fast internal capture process of Fig. \ref{f4}.  
For models without the internal capture, the main mechanism for \bes\ depletion is the 
catalysis of $p$-burning of \bex\ bound states that at $0.2\la T_9 \la 0.3$ is faster than the 
Hubble rate. At the same time, we note that the problem of calculating \bes+\lisv\ abundance in CBBN is 
in certain ways harder than the previously considered case of \lisx, simply because it requires better than 30\% accuracy
for the answer, while a factor of a few error in \lisx\ can still be tolerated given the uncertain 
observational status of \lisx.

The allowed part of the parameter space requires $\tau_X$ to be less than about 2000 seconds, and the 
initial abundance to be of order 0.05 or larger. Interestingly enough, a very similar range of lifetimes was suggeted in 
\cite{jj} where the lithium overproduction problem is solved via the hadronic energy release. 
Even though such abundances can be achieved in particle physics 
models, it is not difficult to see that they require the total energy density carried by \xm\ 
be equal or {\em larger} than the energy density of cold dark matter. Assuming that 
the minimal allowed mass for \xm\ is on the order of 100 GeV, we get the following estimate for the 
energy density of \xm\, before their decay
\be
\frac{\Omega_{X^-}}{\Omega_b} \simeq \fr{Y_X m_X}{m_p} \geq 5.
\ee
Since it is natural to expect that the same amount of energy would be concentrated in \xp, 
we come to the conclusion that in order to solve lithium problem \xm\ and \xp\ would have to carry 
a factor of two more energy than the cold dark matter. This forces a conclusion that the decays of 
charged $X$ particles in these models would have to be accompanied by a significant energy release. 
There is, however, a caveat of the nuclear uncertainty in the recombination rate of \bes\ and 
\xm, and only more elborate nuclear calculations can clarify this picture.

It is important to realize that the efficiency of the CBBN reduction of \lisv+\bes\ is one-to-one related to 
the photorecombination of \bex, or in other words \bex\ serves as a ``bottleneck" for the CBBN reduction of 
lithium, much like D formation is the bottleneck for the helium formation in SBBN. 
The calculation of the  non-resonant contribution of \bes\ capture directly to the ground state of (\bes\xm) does
not requires significant nuclear physics input, and we believe that the current paper provides a reasonable,
perhaps $\sim 30\%$, accuracy for that part of the cross section. The impact of internal 
excitations of \bes\ in the process of capture may lead to the additional resonant contributions. 
In particular, we find that the $2s$ atomic resonance between the spin 1/2 (excited) state of \bes\ and 
\xm\ is very close to the threshold, and may lead to the factor of a few enhancement in the capture rate. 
Unfortunately, we are unable at this point to improve the calculation to the point when the 
contribution of this resonance is firmly established. In practice, one needs to calculate 
the binding energies in the excited states with $\sim 10$keV accuracy, which is possible only 
with the advanced techniques of modern nuclear physics.
Therefore, such calculation would perhaps be the most important 
rate calculation for the whole CBBN paradigm. 

\begin{figure}[htbp]
\centerline{\includegraphics[width=13 cm]{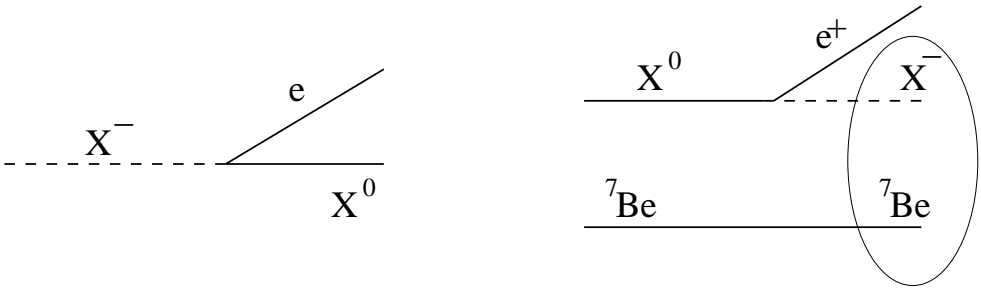}}
 \caption{\footnotesize If the splitting between \xm\ and \xz\ is between 0.511 and 0.822 MeV, 
both the decay of \xm\ to \xz\ and the \xz\ capture by \bes\ become possible. At the same time 
the (\hef\xm) bound state cannot be formed from \xz\ and \hef\ and the \lisx\ overproduction is automatically 
avoided if the lifetime of \xz\ is arbitrarily short. }
\label{f7} 
\end{figure}

There is one alternative CBBN scenario that we would like to briefly 
mention here. Suppose there is a pair of closely degenerate massive $X^\pm$ particles 
and stable dark matter $X^0$ particles 
that can form a vertex together 
with an electron line, Fig. \ref{f7}. 
This can happen if $X^\pm$ are fermions and $X^0$ are bosons or vice versa. 
If the mass difference of the \xm\ and \xz\ satisfies the following relation,
\be
m_e < m_{X^-} - m_{X^0} < 1.33{\rm MeV} - m_e 
\ee 
the two processes are allowed: the decay of \xm\ and \xp\ to \xz, and \xz$-$\bes\ 
recombination,
\be
X^\pm \to X^0+ e^\pm;~~~ X^0+ \besm \to (\besm X^-) +e^+
\label{X0capture}
\ee
In this case, the catalytic destruction of 
\bes\ will proceed the same way as in type I models. Since the coupling between $e$, \xm\ and \xz\ is in 
principle a free parameter, one could tune it to sufficiently large values so that the 
second process in (\ref{X0capture}) exceeds the rate in (\ref{sigma_tot}), thus enabling more efficient 
catalytic suppression of \lisv+\bes. This model would predict dark matter capture by 
nuclei continuing now, which is an intriguing option to consider.  
Detailed analysis of this model falls outside the scope of 
this paper. 

To conclude, this paper provides the detailed account of main processes responsible for the catalytic 
suppression of \lisv+\bes\ in CBBN. The efficiency of such suppression depends mostly on the 
recombination rate leading to \bex, that is calculated in detail. This paper treats only the 
change in the thermal processes, although the non-thermal effects due to the decay and annihilation 
of charged relic particles are also expected to be important. Thus the question of whether or not 
the lithium abundance can be sufficiently depleted by the combination of CBBN processes and the 
energy release should be addressed on model-to-model basis. The 
stau-NLSP/gravitino-LSP model has already been extensively studied in the literature, 
\cite{stau1} and \cite{stau2}. 
Subsequent work  \cite{future} will address the issue of 
whether the significant depletion of \lisv+\bes\ can be achieved in the 
constrained MSSM models with stau-NLSP/neutralino-LSP scenario, when the 
decay of staus is delayed due to the small mass difference with neutralino.

{\bf Acknowledgments} 
M.P. would like to thank R. Cyburt for helpful discussions. Research at the Perimeter Institute 
is supported in part by the Government
of Canada through NSERC and by the Province of Ontario through MEDT.


\begin{thebibliography}{99}

\bi{WMAP} D.~N.~Spergel {\it et al.}  [WMAP Collaboration],
  Astrophys.\ J.\ Suppl.\  {\bf 148}, 175 (2003).
  
  \bi{Sarkar} For a review of BBN constraints see {\em e.g.}: S.~Sarkar,
  Rept.\ Prog.\ Phys.\  {\bf 59}, 1493 (1996).
  
  \bi{Ed} E.~J.~Copeland, M.~Sami and S.~Tsujikawa,
  Int.\ J.\ Mod.\ Phys.\  D {\bf 15}, 1753 (2006)
  [arXiv:hep-th/0603057].


  \bi{metastable} D.~Lindley,
  Astrophys.\ J.\  {\bf 294} (1985) 1; 
  J.~R.~Ellis, D.~V.~Nanopoulos and S.~Sarkar,
  Nucl.\ Phys.\ B {\bf 259}, 175 (1985);
  R.~J.~Scherrer and M.~S.~Turner,
  Phys.\ Rev.\ D {\bf 33}, 1585 (1986);
  M.~H.~Reno and D.~Seckel,
  Phys.\ Rev.\ D {\bf 37}, 3441 (1988);
  S.~Dimopoulos {\em et al.},
  Nucl.\ Phys.\ B {\bf 311}, 699 (1989);
  Yu. L. Levitan {\em et al},  Yad. Fiz. {\bf 47}, 168
(1988) 
[Sov. J. Nucl. Phys. {\bf 47}, 109 (1988)];
R.~H.~Cyburt, J.~R.~Ellis, B.~D.~Fields and K.~A.~Olive,
  Phys.\ Rev.\ D {\bf 67}, 103521 (2003);
 M.~Kawasaki, K.~Kohri and T.~Moroi,
  Phys.\ Rev.\ D {\bf 71}, 083502 (2005);
  K.~Jedamzik,
  Phys.\ Rev.\  D {\bf 74}, 103509 (2006)
  [arXiv:hep-ph/0604251];   M.~Kusakabe, T.~Kajino and G.~J.~Mathews,
  Phys.\ Rev.\  D {\bf 74}, 023526 (2006)
  [arXiv:astro-ph/0605255].

  
  \bibitem{inhom}
  G.~M.~Fuller, G.~J.~Mathews and C.~R.~Alcock,
  Phys.\ Rev.\ D {\bf 37}, 1380 (1988).
  
  \bibitem{alpha} E.~W.~Kolb, M.~J.~Perry and T.~P.~Walker,
  Phys.\ Rev.\ D {\bf 33} (1986) 869; B.~A.~Campbell and K.~A.~Olive,
  Phys.\ Lett.\  B {\bf 345}, 429 (1995)
  [arXiv:hep-ph/9411272].
 
 
\bi{CBBN}  M.~Pospelov,
  Phys.\ Rev.\ Lett.\  {\bf 98}, 231301 (2007)
  [arXiv:hep-ph/0605215].

 
 
\bi{french} A.~Coc {\em et al.},
  Astrophys.\ J.\  {\bf 600}, 544 (2004);  

\bi{belg} C.~Angulo {\it et al.},
  Astrophys.\ J.\  {\bf 630}, L105 (2005)
  [arXiv:astro-ph/0508454].
 
  \bi{astro}  M.~Asplund {\em et al.},
  Astrophys.\ J.\  {\bf 644}, 229 (2006)
  [arXiv:astro-ph/0510636]; 
Korn A.J. {\em et al.},  
Nature 442, 657 (2006); V.~Tatischeff and J.~P.~Thibaud,
  arXiv:astro-ph/0610756; 

  
   
\bi{jj} K.~Jedamzik,
  Phys.\ Rev.\ D {\bf 70}, 063524 (2004); K.~Jedamzik {\em et al.},
  arXiv:hep-ph/0512044.
  
\bi{Khlopov} K.~Belotsky {\em at al.},
  arXiv:hep-ph/0411271; D.~Fargion, M.~Khlopov and C.~A.~Stephan,
  arXiv:astro-ph/0511789; K.~M.~Belotsky, M.~Y.~Khlopov and K.~I.~Shibaev,
  arXiv:astro-ph/0604518. 
  
  \bi{KT} K.~Kohri and F.~Takayama,
  arXiv:hep-ph/0605243.


\bi{KR} M.~Kaplinghat and A.~Rajaraman,
  Phys.\ Rev.\ D {\bf 74}, 103004 (2006)
  [arXiv:astro-ph/0606209].
  
  

\bi{Cy} R.~H.~Cyburt {\em et al.},
  JCAP {\bf 0611}, 014 (2006)
  [arXiv:astro-ph/0608562].
  
  
   
  \bibitem{Dimopoulos:1989hk}
  S.~Dimopoulos, D.~Eichler, R.~Esmailzadeh and G.~D.~Starkman,
  Phys.\ Rev.\  D {\bf 41}, 2388 (1990).

  
 
\bi{radii} I. Tamihata {\em et al.}, Phys.\ Lett.\ B {\bf 206}, 592 (1988).

\bibitem{Baye}
  D.~Baye,
  Phys.\ Rev.\ C {\bf 62}, 065803 (2000).
  
  \bibitem{AD} A. Adahchour and P. Descouvemont, J. Phys. G: Nucl. Part. Phys. {\bf 29} 395, (2003).    

\bi{annih} K.~Jedamzik,
  Phys.\ Rev.\ D {\bf 70}, 083510 (2004)
  [arXiv:astro-ph/0405583].

\bi{Keith} R.~H.~Cyburt, B.~D.~Fields and K.~A.~Olive,
  Phys.\ Lett.\  B {\bf 567}, 227 (2003)
  [arXiv:astro-ph/0302431].

 
\bi{Rich}  R.~H.~Cyburt,
  Phys.\ Rev.\  D {\bf 70}, 023505 (2004)
  [arXiv:astro-ph/0401091].


\bi{reactions} C. Angulo et al., Nucl. Phys. A656 (1999) 3.
 
\bi{future} M. Pospelov and Y. Santoso, {\em in progress}. 


 \bi{newCBBN}  K.~Hamaguchi, T.~Hatsuda, M.~Kamimura, Y.~Kino and T.~T.~Yanagida,
  Phys.\ Lett.\  B {\bf 650}, 268 (2007).

  
\bi{stau1} J.~L.~Feng, A.~Rajaraman and F.~Takayama,
  Phys.\ Rev.\ Lett.\  {\bf 91}, 011302 (2003) and 
  Phys.\ Rev.\  D {\bf 68}, 063504 (2003); J.~L.~Feng, S.~F.~Su and F.~Takayama,
  Phys.\ Rev.\  D {\bf 70}, 063514 (2004) and Phys.\ Rev.\  D {\bf 70}, 075019 (2004).

\bi{stau2} J.~R.~Ellis, K.~A.~Olive, Y.~Santoso and V.~C.~Spanos,
  Phys.\ Lett.\  B {\bf 588}, 7 (2004); W.~Buchmuller, K.~Hamaguchi, M.~Ratz and T.~Yanagida,
  Phys.\ Lett.\  B {\bf 588}, 90 (2004); W.~Buchmuller, K.~Hamaguchi, M.~Ibe and T.~T.~Yanagida,
  Phys.\ Lett.\  B {\bf 643}, 124 (2006);
   F.~D.~Steffen,
  JCAP {\bf 0609}, 001 (2006);
J.~Pradler and F.~D.~Steffen,
  Phys.\ Lett.\  B {\bf 648}, 224 (2007).


\end{thebibliography}
\end{document}